\documentclass[superscriptaddress,aps,prl,reprint,twocolumn,amsmath,amssymb]{revtex4-2}
\usepackage{graphicx}
\usepackage{hyperref}
\usepackage{amssymb}
\usepackage{slashed}
\usepackage{dcolumn}
\usepackage{amsmath}
\usepackage{bm}
\usepackage{colordvi}
\usepackage{algorithm}
\usepackage{algpseudocode}
\usepackage{multirow}
\usepackage{titlesec}
\usepackage{newtxtext,newtxmath}
\usepackage{dsfont}
\usepackage{xcolor}
\usepackage{mathbbol}

\usepackage{hyperref}
\hypersetup{
    colorlinks=true,
    linkcolor=blue,
    citecolor=blue,     
    urlcolor=blue,
}

\allowdisplaybreaks

\usepackage{mathrsfs}
\makeatletter

\newcommand{\Rmnum}[1]{\expandafter\@slowromancap\romannumeral #1@}
\makeatother

\begin{document}

\title{Superconducting Decoherence and Thermal Quenching of the Josephson Diode Effect in Low-Dimensional Josephson Systems}

\author{F. Yang}
\email{fzy5099@psu.edu}

\affiliation{Department of Materials Science and Engineering and Materials Research Institute, The Pennsylvania State University, University Park, PA 16802, USA}

\author{C. Y. Dong}

\affiliation{Department of Materials Science and Engineering and Materials Research Institute, The Pennsylvania State University, University Park, PA 16802, USA}

\author{Joshua A. Robinson}

\affiliation{Department of Materials Science and Engineering and Materials Research Institute, The Pennsylvania State University, University Park, PA 16802, USA}

\author{L. Q. Chen}
\email{lqc3@psu.edu}

\affiliation{Department of Materials Science and Engineering and Materials Research Institute, The Pennsylvania State University, University Park, PA 16802, USA}

\date{\today}

\begin{abstract} 

 Motivated by recent studies on superconducting (SC) diode nonreciprocity,  we uncover a generic smooth SC-phase decoherence mechanism in low-dimensional Josephson structures. Contrary to the conventional single-energy-scale paradigm where Josephson coherence and diode nonreciprocity vanish simultaneously only at the SC gap-closing temperature, 
 we demonstrate, within a fully self-consistent microscopic framework beyond mean-field theory, that SC phase fluctuations generically split these phenomena into distinct energy scales. As a result, rather than a single SC-normal transition, the system exhibits a sequence of distinct thermal crossovers upon heating:  the diode effect disappears first at $T_{\eta}$, Josephson coherence is subsequently lost at $T_c$, and the SC gap collapses only at a higher temperature $T_s$.  Using a bilayer SC system as a concrete example, we show that the separation between these temperature scales is not solely dictated by Josephson coupling, but is instead strongly and counterintuitively shaped by the  in-plane disorder and carrier density. These findings reveal that smooth
SC phase decoherence introduces a distinct and more fragile energy scale, with potential implications for layered superconductors such as cuprates and recently discovered nickelates, as well as for SC qubit platforms.

\end{abstract}

\maketitle  

\begin{figure}[htb]
  \includegraphics[width=8.0cm]{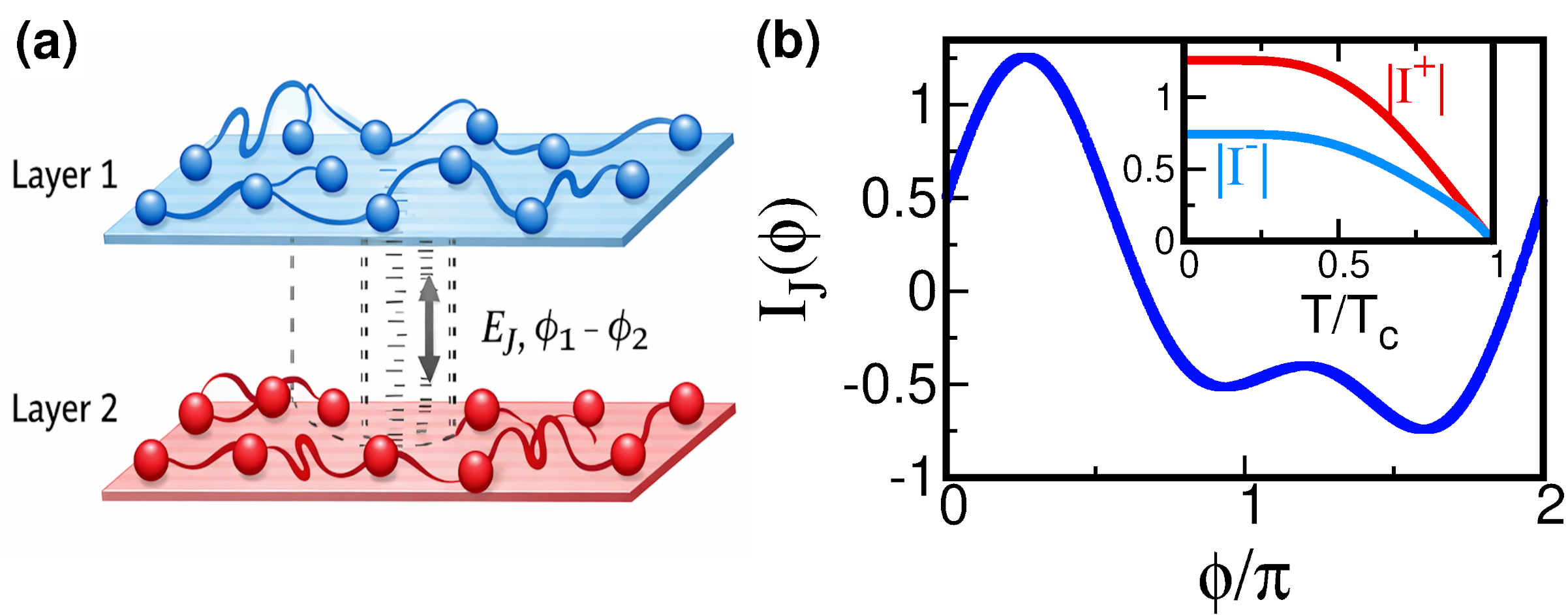}
  \caption{
  ({\bf a}) Schematic illustration of a low-dimensional Josephson system. 
  ({\bf b}) The CPR given by Eq.~(\ref{CPR}).
  Inset: Critical-current strength (normalized) in the forward and backward directions from Eq.~(\ref{CPR}) based on the mean-field BCS theory to calculate SC gap. Without loss of generality, we set $\alpha=0.2\pi$, $|\Delta_1|=|\Delta_2|$, and $|J_1|=4|J_2||\Delta_1||\Delta_2|$.}
  \label{figyc1}
\end{figure}

{\sl Introduction.---}Over the past few decades, Josephson junctions, where two superconductors are coupled through a non-superconducting barrier~\cite{RevModPhys.46.251}, have emerged not only as the basis of a wide range of superconducting (SC) electronic devices~\cite{Bal2012,Wallraff2004,doi:10.1126/science.abf5539}, but also as essential building blocks of quantum technologies and quantum computation~\cite{Ioffe1999,PhysRevLett.89.197902,doi:10.1126/science.1069452}. 
The physics of Josephson systems is governed by the current-phase relation (CPR) of the Josephson supercurrent $I_J$ as a function of the SC phase difference $\phi$ between the two electrodes~\cite{RevModPhys.76.411}, which acquires a generalized form~\cite{Pal2022,PhysRevB.109.094518,Can2021}  upon a standard perturbative treatment~\cite{mahan2013many,PhysRevLett.118.016802} of the electron tunneling amplitude:
\begin{equation}
I_J(\phi)=|J_1||\Delta_1||\Delta_2|\sin(\phi+\alpha)
+2|J_2||\Delta_1|^2|\Delta_2|^2\sin(2\phi).
\label{CPR}
\end{equation}
Here $|J_1|$ and $|J_2|$ denote the amplitudes of the first and second harmonics, and $|\Delta_{1,2}|$ are the SC gaps of the two electrodes. 
The phase shift $\alpha$ is a gauge-invariant quantity that cannot be removed by a gauge transformation. 
In conventional Josephson junctions with a classical or centrosymmetric barrier, $\alpha=0$, and the CPR is symmetric, leading to identical critical supercurrents for opposite current directions and therefore reciprocal SC  transport. In contrast, $\alpha\ne0$ signals the absence of time-reversal symmetry~\cite{PhysRevX.12.041013}, typically induced by internal or external magnetism, resulting in an asymmetric CPR [as shown in Fig.~\ref{figyc1}(b)].  This asymmetry gives rise to the Josephson diode effect, namely a nonreciprocal critical supercurrent. 

By analogy with semiconductor $p$–$n$ junctions that allow low-resistance charge transport in one direction but high resistance in the opposite direction, Josephson diodes enable directional SC transport, and has therefore attracted considerable attention recently~\cite{PhysRevLett.128.037001,PhysRevLett.128.177001,YuanFu2022,PhysRevLett.129.267702,
doi:10.1126/sciadv.abo0309,Pal2022,rqp1-jtcb,PhysRevLett.130.177002,
doi:10.1126/sciadv.adw6925,Anwar2023}. Owing to  the rapid development of advanced quantum materials, the Josephson diode behavior has been experimentally and theoretically explored in low-dimensional junctions, including a variety of multilayer and bilayer platforms~\cite{wei2025scalable,Ghosh2024,Wu2022,PhysRevLett.130.266003,DiezMerida2023,ccb4-tqxq,Qi2025,Wang2026,Narita2022,Can2021}. Such Josephson junctions can host engineered asymmetry, for example arising from twist angles~\cite{Ghosh2024,DiezMerida2023,ccb4-tqxq,Qi2025,Wang2026,Can2021,PhysRevLett.130.266003}, offering a versatile and highly tunable route toward controlling the Josephson diode behavior. 

Within the conventional paradigm, all Josephson phenomena, including phase coherence and diode nonreciprocity, are expected to disappear only when the SC gap collapses, even in low dimensions, since the Berezinskii-Kosterlitz-Thouless (BKT) mechanism, the standard mechanism for destroying SC phase coherence through vortex unbinding, does not apply to the Josephson phase that is locked by  Josephson couplings. Here we demonstrate, within a self-consistent microscopic framework beyond mean-field theory, that smooth SC  phase decoherence generically introduces a distinct and more fragile energy scale in low-dimensional Josephson systems such as bilayer superconductors. Consequently, Josephson phase coherence and hence all Josephson effects disappears well before the SC gap collapses. Strikingly, the Josephson diode nonreciprocity vanishes at an even lower temperature, preceding the loss of Josephson coherence itself. This reveals a generic SC-decoherence mechanism in low-dimensional Josephson systems, distinct from the BKT mechanism, and instead of a single SC-normal transition, the system undergoes a sequence of thermal crossovers upon heating  prior to the SC-gap collapse.

{\sl Model.---}We start with an effective Hamiltonian description~\cite{abrikosov2012methods,schrieffer1964theory}, taking a bilayer superconductor with interlayer Josephson coupling as an example [as illustrated in Fig.~\ref{figyc1}(a)], 
\begin{align}
H=&\!\!\sum_{j=1,2}\Big[\!\int\!{d{\bf x}}\,\Psi^{\dagger}_{j}({\bf x})\left(\begin{array}{cc}
\xi_{j{\hat {\bf p}}} & \Delta_j({\bf x}) \\
\Delta_j^*({\bf x})& \xi_{j{\bf {\hat p}}} 
\end{array}\right)\Psi_{j}({\bf x})\!-\!\frac{|\Delta_j|^2}{U}\Big]\nonumber\\
&\mbox{}-\frac{1}{2}(J_1\Delta_{1}\Delta^*_{2}\!+\!c.c.)\!-\!\frac{1}{2}[J_2(\Delta_{1}\Delta^*_{2})^2\!+\!c.c.]\!+\!H_I,
\end{align}
where $\Psi^{\dagger}_{j}=(c^{\dagger}_{j\uparrow},c_{j{\downarrow}})$ is the Nambu spinor in layer $j$ with $c^{\dagger}_{js}$ and $c_{js}$ being the creation and annihilation operators, respectively; $\xi_{j{\bf k}}$ denotes the single-particle dispersion measured from the chemical potential, and $U<0$ is the SC pairing interaction; the momentum operator $\hat{\bf p}=-i\hbar{\nabla}$, while ${\bf x}$  denotes the spatial coordinate  confined to the in-plane space of each layer; $H_I$ denotes the long-range Coulomb interactions (see Sec.~SI). The SC order parameter in each layer reads~\cite{abrikosov2012methods,schrieffer1964theory}
\begin{equation}
\Delta_j({\bf x})=|\Delta_j|e^{i\phi_j({\bf x})},
\end{equation}
where $|\Delta_j|$ and $\phi_j({\bf x})$ are the SC gap and phase, respectively. We consider the symmetric bilayer case and assume identical electronic structures in the two layers, i.e., $\xi_{j{\bf k}}=\xi_{\bf k}$, thereby yielding $|\Delta_j|=|\Delta|$. The interlayer Josephson coupling here includes both the conventional first harmonic and a higher-order second harmonic contribution~\cite{Pal2022,PhysRevB.109.094518,Can2021}. Within the conventional derivation of the Josephson effect ~\cite{mahan2013many,PhysRevLett.118.016802}, these couplings originate from the interlayer hopping of electrons, which mediates the Cooper-pairs tunneling between the two SC layers via virtual intermediate states. The first harmonic $J_1$ arises from the lowest-order tunneling of a single Cooper pair, and the second harmonic $J_2$ emerges from correlated tunneling processes of two Cooper pairs. These Josephson amplitudes are generally complex. By gauge transformation, one can take a complex $J_1=|J_1|e^{i\alpha}$, while taking $J_2$ to be real, $J_2=J_2^*$.  The phase shift $\alpha$ represents a nontrivial relative phase between different Josephson harmonics that explicitly breaks time-reversal symmetry, such that 
 $2{\rm arg}(J_1)\ne{\rm arg}(J_2)$, and cannot be removed by a global redefinition of the Josephson  phases.  Such a phase shift can arise from the interplay of interlayer twist effect~\cite{PhysRevX.12.041013,YuanFu2022,doi:10.1126/sciadv.abo0309,Ghosh2024,DiezMerida2023,ccb4-tqxq,Qi2025,Wang2026,Can2021,PhysRevB.109.094518,PhysRevLett.130.266003}, which generates multiple inequivalent tunneling channels, and external or internally generated effective magnetic fields that imprint phase frustration onto these channels.

To consider the SC phase coherence, we introduce the relative (Josephson) and total SC phase fields as
\begin{eqnarray}
\phi({\bf x})&=&\phi_1({\bf x})-\phi_2({\bf x}),\\
\theta({\bf x})&=&[\phi_1({\bf x})+\phi_2({\bf x})]/2.
\end{eqnarray}
Within path-integral formalism, integrating out the fermionic degrees of freedom leads to an effective action (Sec.~SI)
\begin{equation}
S_{\rm eff}= S_{\rm intra}\big(|\Delta|,{\bf p}_s^{2}\big) + S_J(\phi).
\end{equation} 
Here, $S_{\rm intra}$ governs the intralayer SC properties, while ${\bf p}_s=\bm{\nabla}\theta({\bf x})/2$ is the physically meaningful, gauge-invariant quantity describing  fluctuations of the total phase
 $\theta({\bf x})$~\cite{benfatto10,yang19gauge},  which correspond to the in-phase collective mode~\cite{sun20collective,yang19gauge,yang21theory}, associated with the bosonic Nambu–Goldstone (NG) mode of the SC state~\cite{nambu2009nobel,nambu1960quasi,goldstone1961field,goldstone1962broken}, and involve the long-range Coulomb interactions in the symmetric channel (see Sec.~SI). 

The microscopically derived effective action for the interlayer Josephson coupling and the out-of-phase collective mode (known as the Leggett-type~\cite{Leggett1966} Josephson phase mode) reads
\begin{align}
S_J(\phi)=&\!\int\!{dR}\Big[\frac{D_l}{2}\Big(\frac{\partial_t\phi}{2}\Big)^2\!-\!\frac{f_s}{4}\Big(\frac{\nabla\phi}{2}\Big)^2\!+\!|J_1||\Delta|^2\cos(\phi\!+\!\alpha)\nonumber\\
&\mbox{}+|J_2||\Delta|^4\cos(2\phi)\Big].
\end{align}
Here, $D_l$ denotes the effective density of states renormalized by the 
long-range Coulomb interactions in the anti-symmetric channel (see Sec.~SI), and $f_s$ is the phase stiffness. Clearly, the interlayer Josephson coupling explicitly gaps the out-of-phase mode. We decompose the Josephson phase as
 $\phi=\phi_e+\delta\phi$ where $\delta\phi$ represents Josephson phase fluctuations around equilibrium configuration $\phi_e$. With $\langle\delta\phi\rangle=0$, using the cumulant expansion (see Sec.~SII.~A for detailed derivations)
\begin{align}\label{DW}
&\langle\cos(m\phi)\rangle\!=\!{\rm Re}[e^{im\phi_e}\langle{e^{im\delta\phi}}\rangle]\!=\!{\rm Re}\Big[e^{im\phi_e}\!\!\sum_{n=0}\frac{\langle(im\delta\phi)^{2n}\rangle}{(2n)!}\Big]\nonumber\\
&={\rm Re}\Big[e^{im\phi_e}\sum_{n=0}\frac{(2n-1)!!(-1)^nm^{2n}\langle\delta\phi^{2}\rangle^n}{(2n)!}\Big]\nonumber\\
&={\rm Re}[e^{im\phi_e}e^{-\langle{\delta\phi}^2\rangle/2}]=\cos(m\phi_e)e^{-m^2\langle{\delta\phi}^2\rangle/2}, 
\end{align}
the thermally averaged Josephson energy reads
\begin{eqnarray}
E_J(\phi_e)
&=&
-|J_1||\Delta|^2
\big\langle \cos(\phi+\alpha) \big\rangle
-|J_2||\Delta|^4
\langle\cos(2\phi)\rangle \nonumber\\
&=&
-{\bar J}_1(T)\cos(\phi_e+\alpha)-
{\bar J}_2(T)
\cos(2\phi_e),\label{EJ}
\end{eqnarray}
where the effective Josephson couplings are renormalized by Josephson phase fluctuations as 
\begin{eqnarray}
{\bar J}_1(T)&=&|J_1||\Delta(T)|^2\exp[-\langle \delta\phi^2(T)\rangle/2],\label{J1} \\
{\bar J}_2(T)&=&|J_2||\Delta(T)|^4\exp[-2\langle \delta\phi^2(T)\rangle]. \label{J2}
\end{eqnarray}
The optimal equilibrium phase  $\phi^{\rm op}_e$  is  determined by global minimizer of the fluctuation-renormalized Josephson energy
 \begin{equation}
 \phi_e^{\rm op}=\arg\min_{\phi^e\in[0,2\pi)}E_J(\phi_e).
 \end{equation}
The Josephson phase fluctuations can be derived within the quantum statistic mechanism based on the equation of motion of Josephson phase fluctuations $\delta\phi$ obtained from the action $S_{J}(\phi)$ (Sec.~SII.~B), and their average is derived as (Sec.~SII.~C)
\begin{equation}\label{JPF}
\langle \delta\phi^2(T)\rangle=\int\frac{2d{\bf q}}{(2\pi)^2}\frac{2n_B[\omega_{\rm L}({q})]+1}{D_l\omega_{\rm L}({q})},
\end{equation}
which corresponds to the standard bosonic excitation, consisting of contributions from both thermal excitations $2n_B(\omega_{\rm L})$ (thermal fluctuations) and zero-point oscillations (quantum fluctuations). Here, the function $n_B(x)$ is the Bose distribution. The excitation spectrum $\omega_{\rm L}(q)$ of this  Leggett-type~\cite{Leggett1966} collective phase mode is determined by 
\begin{equation}\label{omega}
D_l\omega^2_{\rm L}({q})={4{\bar J}_1\cos(\phi^{\rm op}_e\!+\!\alpha)}+{16{\bar J}_2\cos(2\phi^{\rm op}_e)}
+{f_sq^2}/{2},
\end{equation}
corresponding to a massive (gapped)  collective excitation.

Consequently, Eqs.~(\ref{EJ})–(\ref{omega}) form a nontrivial closed set of self-consistent equations, which  determine the effective Josephson couplings ${\bar J}_1(T)$ and ${\bar J}_2(T)$ renormalized by the Josephson phase fluctuations. Further considering the Josephson supercurrent injection $I_J$ (Sec.~SII.~A), this procedure  yields a qualitatively modified Josephson CPR, 
\begin{equation}
I_J(\phi_e)=2e[{\bar J}_1(T)\sin(\phi_e+\alpha)+2{\bar J}_2(T)\sin(2\phi_e)].
\end{equation}
The Debye-Waller-like factors appearing in the Josephson couplings, $\exp({-\langle{\delta\phi^2}(T)\rangle}/2)$ in ${\bar J}_1(T)$ and $\exp({-2\langle{\delta\phi^2}(T)\rangle})$ in ${\bar J}_2(T)$, which originate from Josephson phase fluctuations, render the Josephson CPR no longer governed solely by a single energy scale set by the SC gap. These factors encode the decoherence of the Josephson coupling, and the resulting Josephson response is governed jointly by the SC gap and phase coherence. It should be emphasized that this decoherence mechanism is fundamentally different from BKT-type transitions: it does not involve vortex unbinding, topological defects of the global phase, or the characteristic universal jump associated with the BKT transition. Instead, it arises from the smooth decoherence of relative-phase locking between SC components. 

In particularly, in the limit where $\bar J_1 \rightarrow 0$ and $\bar J_2 \rightarrow 0$ as a consequence of signigicant  Josephson phase fluctuations, while the SC gap $|\Delta|$ remains finite, the equilibrium Josephson phase can no longer be uniquely determined from the minimum of the Josephson energy $E_J$ in Eq.~(\ref{EJ}). This indicates that SC phase coherence across the Josephson junction is not established, despite the presence of a finite SC pairing gap.

{\sl Results.---}We perform fully self-consistent numerical calculations of the proposed  framework [Eqs.~(\ref{EJ})–(\ref{omega})].  The phase stiffness $f_s$ is written as (see Sec.~SI for detailed derivation)
\begin{equation}\label{fss}
f_s=\!\frac{\mathrm{v}^2_{F}|\Delta|^2}{1+\xi/l}\sum_{\bf k}\frac{f(E_{\bf k}^-)-f(E_{\bf k}^+)}{2E^3_{\bf k}},
\end{equation}
which depends on gap $|\Delta(T)|$ and  NG phase fluctuations $\langle {\bf p}_s^{2} \rangle$.  Here, the fermionic quasiparticle spectrum $E_{\bf k}^{\pm}={\bf v_{\bf k}}\cdot{\bf p}_{s}\pm{E_{\bf k}}$ with $E_{\bf k}=\sqrt{\xi_{\bf k}^2+|\Delta|^2}$; the prefactor $(1+\xi/l)^{-1}$ accounts for the disorder-induced reduction of the phase stiffness, where $\xi=\hbar v_F/|\Delta|$ is the SC coherence length and $l=v_F\tau$ is the mean free path with $\tau$ being the effective scattering time. Within a parabolic-band approximation, $f_s$ reduces to the familiar form ${n_s}/m^*$~\cite{sun20collective,yang19gauge,yang21theory,PhysRevB.106.144509,PhysRevB.98.094507}, with $n_s$ being superfluid density and $m^*$ the effective mass. In 2D systems, the NG mode remains gapless and an active excitation~\cite{yang21theory,PhysRevB.97.054510,yang2025tractable,yang2025efficient,yang2025preformed}, in contrast to the dynamically inert case in 3D~\cite{ambegaokar61electromagnetic,littlewood81gauge}, where the NG mode is fully gapped out by the Anderson–Higgs mechanism~\cite{anderson63plasmons}. Owing to this active nature, the SC gap and its disappearance temperature $T_s$ acquire nontrivial 
dependence on disorder and carrier density, reflecting fluctuation-induced renormalization effects beyond mean-field theory~\cite{yang2025tractable,yang2025efficient,yang2025preformed}. For completeness, we adopt the developed approach in Refs.~\cite{yang2025tractable,yang2025efficient,yang2025preformed} for the in-plane sector to determine $|\Delta(T)|$ and $\langle {\bf p}_s^{2} \rangle$ (see Sec.~SIII), which enter the phase stiffness for interlayer Josephson formulation. The specific  model parameters used in simulation are provided in the Supplemental Material (Sec.~SIV).

\begin{figure}[H]
  \includegraphics[width=8.7cm]{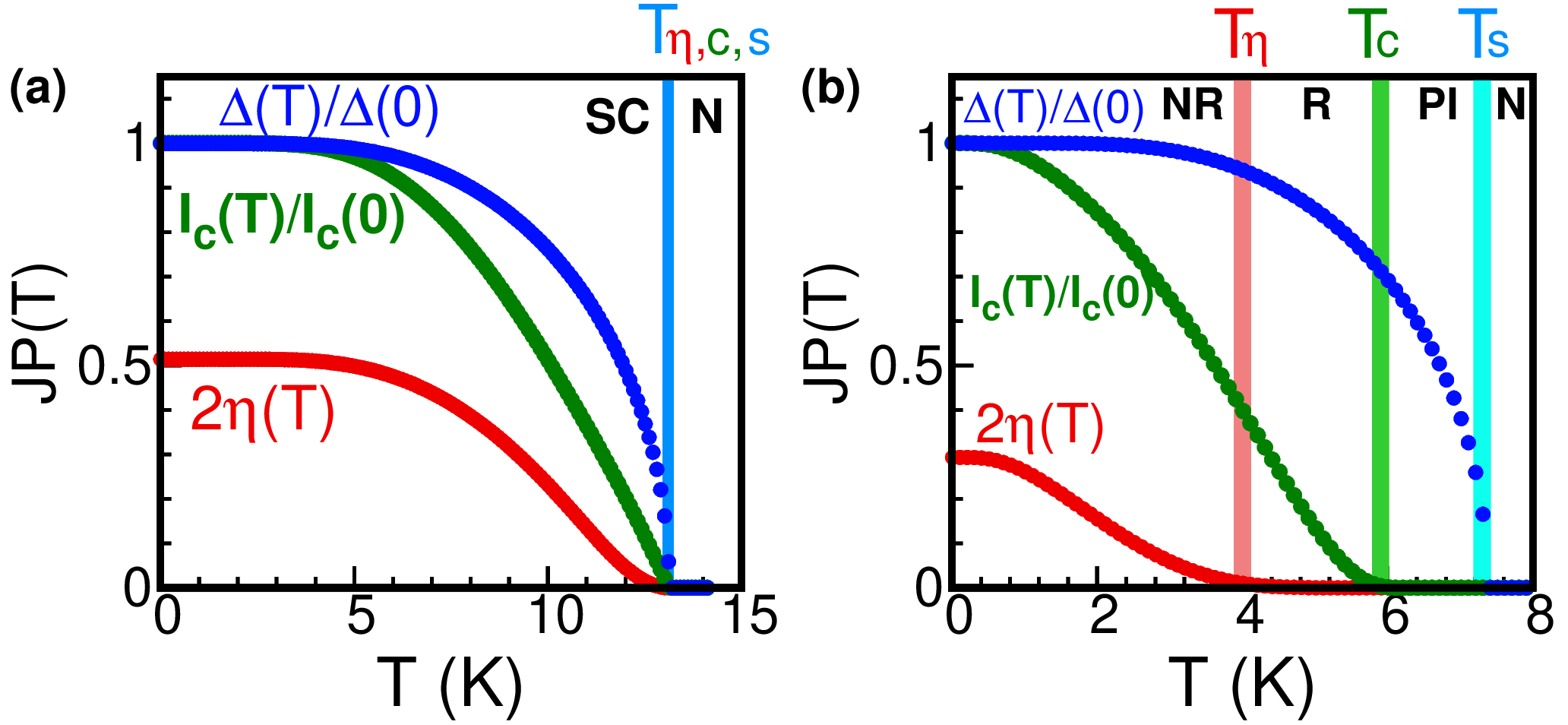}
  \caption{Temperature dependence of the  Josephson critical current, the diode efficiency and the SC gap calculated by ({\bf a}) mean-field and ({\bf b}) phase-fluctuation theories. We set $v_F=10^5~$m/s,  $l=3.3~$nm,  and a BCS constant of $0.55$.  The phase-fluctuation results in ({\bf b}) reveal a sequence of distinct regimes upon heating: rather than a single SC-normal transition predicted by the mean-field theory in ({\bf a}), the system upon heating first evolves from a nonreciprocal (NR) Josephson phase to a reciprocal (R) Josephson phase at $T_{\eta}$, followed by a phase-incoherent (PI) Josephson state at $T_c$, above which phase
coherence across the Josephson junction is lost while the SC gap remains finite, and finally enters the normal (N) state at $T_s$.
 }
  \label{figyc2}
\end{figure}

Following previous works~\cite{Pal2022,PhysRevB.109.094518,Can2021}, we define the Josephson critical current as $I_c(T)={\rm max}(|I_c^+(T)|,|I_c^-(T)|)$, and introduce the diode efficiency: 
\begin{equation}
\eta(T)=\Bigg|\frac{|I_c^+(T)|-|I_c^-(T)|}{|I_c^+(T=0)|+|I_c^-(T=0)|}\Bigg|
\end{equation}
which quantifies the relative strength of the nonreciprocal Josephson response at finite  temperatures, normalized to its $T=0$ value. Here, $I_c^+(T)$ and $I_c^-(T)$ denote the critical Josephson currents for positive and negative current directions, respectively. For comparison, results obtained with and without phase fluctuations are shown in Fig.~\ref{figyc2}.  It is noted that even at $T=0$, 
including the contribution from the zero-point Josephson phase fluctuations suppresses the zero-temperature diode efficiency $\eta(0)$, while the NG and Josephson phase fluctuations also reduce the zero-temperature gap $|\Delta(0)|$ (as discussed in Refs.~\cite{yang2025tractable,yang2025efficient,yang2025preformed}) and the critical current $I_c(0)$ (see Sec.~SV for more details related to zero-point oscillations), respectively. This zero-point effect suggests the intrinsic quantum sensitivity of low-dimensional Josephson-diode systems, and it represents a fundamental quantum limitation that cannot be circumvented by cooling. Quantum fluctuations therefore impose an intrinsic limit on Josephson coherence and nonreciprocal transport in low-dimensional Josephson systems.

For the essential critical behaviors, in the absence of phase fluctuations, as shown in Fig.~\ref{figyc2}(a), both SC and Josephson properties are solely determined by the SC gap, such that the Josephson critical current $I_c(T)$ and the diode efficiency $\eta(T)$ vanish at the same temperature at which the SC gap $|\Delta(T)|$ closes. In contrast, when significant Josephson phase fluctuations are activated, as shown in Fig.~\ref{figyc2}(b),  a clear separation of energy scales emerges upon increasing temperature:
the diode efficiency $\eta(T)$ disappears first, followed by the suppression of the Josephson critical current $I_c(T)$, while the SC gap $|\Delta(T)|$ remains finite and vanishes only at a higher temperature. 

This is because that when Josephson phase fluctuations are activated, the Debye–Waller–like factors in ${\bar J}_1$ and ${\bar J}_2$ lead to a rapid suppression of the effective Josephson couplings as the fluctuation strength increases with temperature. As a result, the Josephson coupling is exponentially reduced by Josephson phase fluctuations, even in a regime where the SC gap remains finite, causing the Josephson effect to vanish at a temperature $T_c$ lower than the gap-closing temperature $T_s$. Importantly, higher-order Josephson couplings are suppressed more strongly than the lowest-order term: the second-harmonic coupling  ${\bar J}_2\propto\exp({-2\langle{\delta\phi^2}\rangle})$  decays parametrically faster than the first-harmonic coupling ${\bar J}_1\propto\exp({-\langle{\delta\phi^2}\rangle}/2)$  as phase fluctuations increase with temperature. 
This hierarchy implies that non-sinusoidal components of the Josephson current–phase relation are particularly fragile against phase decoherence, with higher-order harmonics (e.g., $n\ge3$) expected to be even more fragile according to Eq.~(\ref{DW}). Since the Josephson diode nonreciprocity heavily relies on these non-sinusoidal components~\cite{Pal2022,PhysRevB.109.094518,Can2021}, the Josephson diode effect is expected to disappear at a temperature $T_{\eta}$ lower than the critical $T_c$ at which the Josephson critical current is suppressed. 

\begin{figure}[htb]
  \includegraphics[width=8.6cm]{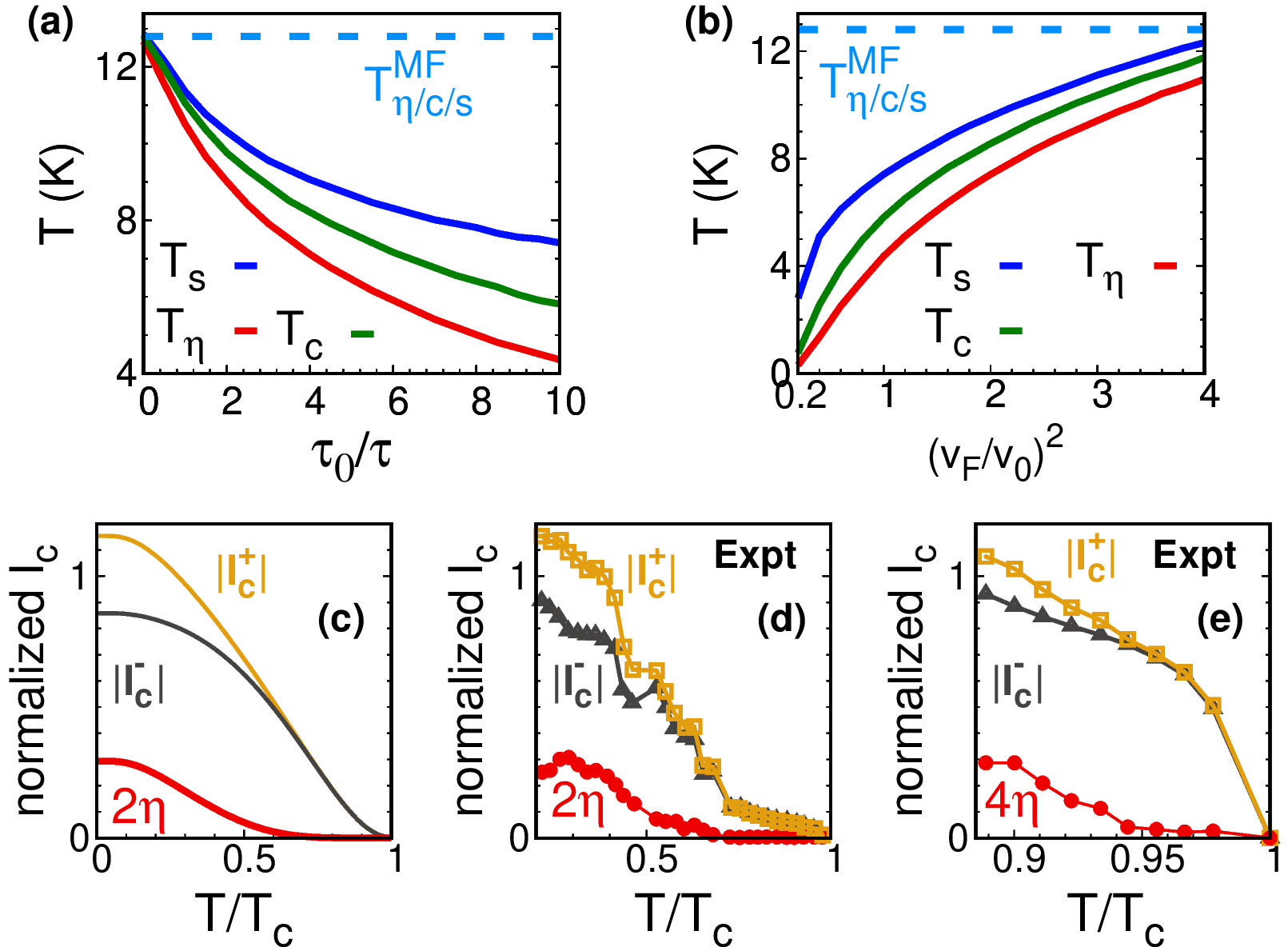}
\caption{Distinct temperatures $T_{\eta}$, $T_c$, and $T_s$ as the phase stiffness is reduced by 
({\bf a}) increasing disorder at $v_F=v_0$ and 
({\bf b}) decreasing the Fermi velocity at $\tau=0.1\tau_0$. Here $v_0=10^5~\mathrm{m/s}$ and $\tau_0v_0=33~$nm. For comparison, we also show the mean-field BCS prediction (dashed curves), where $T_s^{\rm MF}=T_{\eta}^{\rm MF}=T_c^{\rm MF}$, which reduces to a constant independent of disorder and carrier density in 2D.  
({\bf c}) The theoretical critical currents $|I_c^+(T)|$ and $|I_c^-(T)|$ for  
$\tau=0.2\tau_0$ and $v_F=v_0$.  For comparison, we present related  experimental data in ({\bf d}) and ({\bf e}), extracted from Refs.~\cite{Anwar2023} and~\cite{wei2025scalable}, respectively. The qualitative merging of the two critical currents with increasing $T$ before $T_c$ in ({\bf d}) and ({\bf e}) (see also Refs.~\cite{Kim2024,ding2025indirect,Ma2025NbSe2Diode,Yan2025InAsDiode}) is consistent with the theoretical prediction. In principle, quantitative agreement with specific experimental systems can be achieved by tuning the Josephson parameters ($J_1$, $J_2$, and $\alpha$) together with the disorder strength or carrier density.}
  \label{figyc3}
\end{figure}

We emphasize that the renormalized Josephson couplings in Eqs.~(\ref{J1}) and~(\ref{J2}) are derived in a mathematically rigorous manner, without reliance on model-specific parameters or physical assumptions, and are therefore generic. Then, the key point is whether Josephson phase fluctuations can be effectively excited. Typically, the Josephson coupling is weak, such that the gap of the Josephson phase mode [$\omega_L(q=0)$ in Eq.~(\ref{omega})] lies below the SC gap, rendering it a thermally active low-energy excitation. Notably, an equally important factor is the phase stiffness $f_s$, which determines the available phase space for bosonic phase fluctuations entering the fluctuation integrals [Eq.~(\ref{JPF})].
 In bulk systems, the phase stiffness is generally large, strongly suppressing phase fluctuations. By contrast, in low-dimensional systems the phase stiffness is significantly reduced, owing both to a smaller Fermi velocity and to the intrinsically enhanced disorder associated with reduced dimensionality~\cite{Saito2016,Qiu2021}.  Increasing disorder and reducing the Fermi velocity, as shown in Figs.~\ref{figyc3}(a)~and~(b), respectively, lead to a suppression of $T_s$, $T_{c}$, $T_{\eta}$, while markedly increasing the separation among these characteristic temperatures. As a result, the hierarchy of SC decoherence scales becomes increasingly pronounced. Among these effects, the suppression of gap closing temperature $T_s$ originates from the enhanced NG phase fluctuations (as discussed in Refs.~\cite{yang2025tractable,yang2025efficient,yang2025preformed}), whereas all the remaining Josephson features are primarily driven by the enhanced Josephson phase fluctuations (i.e., decoherence).
 
{\it Discussion.—}Beyond the conventional single-energy-scale paradigm of the Josephson physics, the present study  demonstrates that  phase decoherence constitutes an independent and more fragile energy scale in low-dimensional Josephson systems, and hence, Josephson physics itself in this case splits into multiple decoherence regimes, including diode nonreciprocity as the most fragile scale. We therefore predict that the forward and backward critical currents in low-dimensional Josephson-diode systems, $|I_c^+(T)|$ and $|I_c^-(T)|$, to merge as temperature increases, before vanishing altogether at $T_c$, as shown in Fig.~\ref{figyc3}(c). A similar tendency appears to be present in a wide range of recent experiments (e.g., Refs.~\cite{wei2025scalable,Anwar2023,Kim2024,ding2025indirect,Ma2025NbSe2Diode,Yan2025InAsDiode}), where the data [see Fig.~\ref{figyc3}(d)~and~(e)] exhibit a temperature evolution qualitatively consistent with  theoretical prediction.

It should be emphasized that the vanishing of the Josephson critical current and the disappearance of the diode effect correspond to exponential thermal crossovers, rather than conventional thermodynamic phase transitions of first or second order. Both the diode efficiency $\eta(T)$ and the Josephson critical current $I_c(T)$  shown in Fig.~\ref{figyc2}(b)  exhibit behaviors with continuous temperature derivatives in the vicinity of their respective disappearance temperatures,  in contrast to the mean-field scenario shown in Fig.~\ref{figyc2}(a), where discontinuities in temperature derivatives are expected. 
 Consequently, these crossovers do not exhibit thermodynamic signatures such as latent heat or a jump in the specific heat, in contrast to the SC gap-closing transition, which is a genuine second-order phase transition~\cite{abrikosov2012methods,schrieffer1964theory}. Such critical behaviors of the Josephson properties thus constitute a key distinguishing feature between mean-field and fluctuation-based descriptions of Josephson systems. The proposed \emph{smooth}-SC-decoherence mechanism follows analytically from the structure of both quantum and thermal phase fluctuations in low dimensions and does not rely on microscopic pairing details, symmetry-specific mechanisms, or fine tuning. It is fundamentally distinct from the BKT transition, the well-known mechanism for destroying SC phase coherence in two dimensions, which acts on the gapless global phase through vortex unbinding. By contrast, the Josephson phase mode is gapped due to interlayer phase locking and therefore does not participate in BKT physics.

The developed theory provides a tractable description of SC decoherence processes. Owing to the structural similarity~\cite{sun20collective,yang2025preformed}, this framework can be potentially extended to layered superconductors with multilayer structures and Josephson coupling $J_c\cos(\theta_{n+1}-\theta_n)$ between different layers (where $\theta_n$ is the SC phase in the $n$-th layer),  such as cuprates~\cite{dagotto94correlated,tom99the,armitage10progress,davis13concepts,damascelli2003angle,keimer15from,bednorz86possible,wu1987superconductivity} and recently discovered nickelates~\cite{Nomura2022,WangLeeGoodge2024,Li2019,Xu2025,Hepting2020,PhysRevB.101.020501}, particularly for their $c$-axis transport behavior. In this context, one can anticipate the disappearance of $c$-axis superconductivity (zero-resistance) while the SC gap persists above a characteristic critical temperature. Moreover, owing to the close analogy in the underlying derivation of the excitations, similar considerations also apply to SC transmon qubits~\cite{girvin2014circuit,PhysRevA.76.042319,
doi:10.1126/science.1231930}. Our results indicate that higher-order Josephson harmonics are increasingly fragile against decoherence. This effect is  expected to be relevant for the stability/coherence of Josephson-based SC qubits, where such higher harmonics play an essential role in determining the qubit spectrum and anharmonicity.

{\it Acknowledgments.---}This work is supported by the US Department of Energy, Office of Science, Basic Energy Sciences, under Award Number DE-SC0020145 as part of Computational Materials Sciences Program. F.Y. and L.Q.C. also appreciate the  generous support from the Donald W. Hamer Foundation through a Hamer Professorship at Penn State.

%
\end{document}


\title{Superconducting Decoherence and Thermal Quenching of the Josephson Diode Effect in Low-Dimensional Josephson Systems (Supplemental Materials)}

\author{F. Yang}
\email{fzy5099@psu.edu}

\affiliation{Department of Materials Science and Engineering and Materials Research Institute, The Pennsylvania State University, University Park, PA 16802, USA}

\author{C. Y. Dong}

\affiliation{Department of Materials Science and Engineering and Materials Research Institute, The Pennsylvania State University, University Park, PA 16802, USA}

\author{Joshua A. Robinson}

\affiliation{Department of Materials Science and Engineering and Materials Research Institute, The Pennsylvania State University, University Park, PA 16802, USA}

\author{L. Q. Chen}
\email{lqc3@psu.edu}

\affiliation{Department of Materials Science and Engineering and Materials Research Institute, The Pennsylvania State University, University Park, PA 16802, USA}

\maketitle 

\section{Derivation of the effective action}
\label{seceac}

In this section, we provide a microscopic derivation of the low-energy effective action for a bilayer
superconductor with interlayer Josephson coupling. By integrating out the fermionic degrees of freedom within the path-integral approach~\cite{schrieffer1964theory,peskin2018introduction}, we obtain a controlled description of the Josephson phase dynamics and the associated collective
fluctuations.

Specifically, based on the Hamiltonian presented in the main text, the action of the model is given by
\begin{align}S=&\int{dx}\sum_{j=1,2}\Big[\sum_{s=\uparrow,\downarrow}\psi_{js}^{\dagger}(x)(i\partial_t-\xi_{j{\hat{\bf p}}})\psi_{js}(x)-\Delta_j(x)\psi^{\dagger}_{j\uparrow}(x)\psi^{\dagger}_{j\downarrow}(x)-\Delta^*_j(x)\psi_{j\downarrow}(x)\psi_{j\uparrow}(x)+\frac{|\Delta_j(x)|^2}{U}\Big]\nonumber\\
&-\frac{1}{2}\int{dxdx'}\sum_{j}V_{jj}(x-x')\rho^{\dagger}_j(x)\rho_j(x)-\frac{1}{2}\int{dxdx'}\sum_{j}V_{12}(x-x')[\rho^{\dagger}_1(x)\rho_2(x)+\rho^{\dagger}_2(x)\rho_1(x)]+S_{\rm JC},~~~~~~
\end{align}  
where the interlayer Josephson-coupling part is~\cite{Pal2022,PhysRevB.109.094518,Can2021}
\begin{equation}
S_{\rm JC}=\int dx\Bigg\{
\frac{1}{2}\Big(J_1 \Delta_{1}\Delta^*_{2}+J_1^*\Delta_2\Delta_1^*\Big)
+\frac{1}{2}\Big[J_2(\Delta_{1}\Delta^*_{2})^2+J_2^*(\Delta_2\Delta_1^*)^2\Big]
\Bigg\}.
\end{equation}
Here, $\psi_{js}(x)$ is the fermionic field in layer
$j$ with spin $s$;  $x=(t,{\bf x})$ denotes the spacetime coordinate (a four-vector in relativistic notation);
$\hat{\bf p}=-i\hbar{\bm \nabla}$ is the momentum operator;  $U<0$ is the attractive pairing interaction. The density operator is
$\rho_j(x)=\sum_s\psi_{js}^{\dagger}(x)\psi_{js}(x)$. For simplicity and without loss of generality, we assume a parabolic dispersion
$\xi_{j\hat{\bf p}}=\hat{\bf p}^2/(2m)-\mu$ (identical in the two layers).
The long-range Coulomb interaction contains both intralayer contributions
$V_{11}(x-x')=V_{22}(x-x')$ and an interlayer term $V_{12}(x-x')$.
In momentum space they take the form~\cite{sun20collective}
\begin{equation}
V_{jj}(q)=\frac{2\pi e^2}{\varepsilon q},\qquad
V_{12}(q)=\frac{2\pi e^2}{\varepsilon q}\,e^{-qd},
\end{equation}
where $d$ is the interlayer separation and $\varepsilon$ denotes the background dielectric constant.

The superconducting order parameter in each layer is parameterized as~\cite{abrikosov2012methods,schrieffer1964theory}
\begin{equation}
\Delta_j(x)=|\Delta_j|e^{i\phi_j(x)},
\end{equation}
where $|\Delta_j|$ and $\phi_j(x)$ denotes the superconducting gap and phase, respectively.

The interlayer Josephson coupling is assumed to contain both the conventional first harmonic and a
higher-order second harmonic contribution, with generally complex amplitudes $J_1$ and $J_2$
\cite{Pal2022,PhysRevB.109.094518,Can2021}.
 Here we do not repeat the microscopic derivation of these terms, which can be obtained from standard tunneling-Hamiltonian treatments and higher-order perturbation theory in the interlayer hopping amplitude~\cite{mahan2013many,PhysRevLett.118.016802}. Notably, in standard microscopic derivations of the Josephson effect
\cite{mahan2013many,PhysRevLett.118.016802}, these terms originate from interlayer electron hopping, which mediates tunneling of
Cooper pairs between the two superconducting layers. The first harmonic $J_1$ corresponds to the
lowest-order tunneling of a single Cooper pair, while the second harmonic $J_2$ naturally appears from
higher-order processes involving correlated tunneling of two Cooper pairs via virtual intermediate states
\cite{mahan2013many}. Here we take this form as a starting point and further examine the implications of phase coherence. 

Then, applying
the Hubbard-Stratonovich transformation for long-range coulomb interactions and using the unitary transformation $\psi_{js}(x)\rightarrow{e^{i\phi_j(x)/2}}\psi_{js}(x)$~\cite{ambegaokar61electromagnetic,yang21theory,yang19gauge,PhysRevB.98.094507}, the action becomes
\begin{align}&S=\int{dx}\Big[\sum_{s=\uparrow,\downarrow}\psi_{1s}^{\dagger}(x)[i\partial_t-\partial_t\phi_1(x)/2-\xi_{{\hat {\bf p}+{\bm \nabla}\phi_1/2}}-\mu_{+}(x)-\mu_-(x)]\psi_{1s}(x)-|\Delta_1|\psi^{\dagger}_{1\uparrow}(x)\psi^{\dagger}_{1\downarrow}(x)-|\Delta_1|\psi_{1\downarrow}(x)\psi_{1\uparrow}(x)+\frac{|\Delta_1|^2}{U}\Big]\nonumber\\
&+\int{dx}\Big[\sum_{s=\uparrow,\downarrow}\psi_{2s}^{\dagger}(x)[i\partial_t-\partial_t\phi_2(x)/2-\xi_{{\hat {\bf p}+{\bm \nabla}\phi_2/2}}-\mu_{+}(x)+\mu_-(x)]\psi_{2s}(x)-|\Delta_2|\psi^{\dagger}_{2\uparrow}(x)\psi^{\dagger}_{2\downarrow}(x)-|\Delta_2|\psi_{2\downarrow}(x)\psi_{2\uparrow}(x)+\frac{|\Delta_2|^2}{U}\Big]\nonumber\\
&+\int{dtd{\bf q}}\frac{2|\mu_+(t,{\bf q})|^2}{V_+(q)}+\int{dtd{\bf q}}\frac{2|\mu_-(t,{\bf q})|^2}{V_-(q)}+S_{\rm JC},~~~~~~
\end{align}  
Here, $\mu_{\pm}(x)$ denotes the Coulomb-interaction-related Hartree field; the symmetric channel $V_{+}(q)=\frac{2\pi e^2}{\varepsilon q}(1+e^{-qd})$ describes the
long-range Coulomb interaction associated with total charge fluctuations,
while the antisymmetric channel $V_{-}(q)=\frac{2\pi e^2}{\varepsilon q}(1-e^{-qd})$ governs interlayer charge transfer
modes~\cite{sun20collective}.

We introduce the relative (Josephson) and total superconducting phase fields as (in analogy with the derivation of the Leggett mode~\cite{Leggett1966})
\begin{eqnarray}
\phi(x)&=&\phi_1(x)-\phi_2(x),\\\theta(x)&=&[\phi_1(x)+\phi_2(x)]/2.
\end{eqnarray}
and, restrict to the symmetric case
\[
|\Delta_1|=|\Delta_2|=|\Delta|,
\]
for which the two layers are identical at the mean-field level and differ
only through phase and scalar-potential fluctuations. In this case, in Nambu space and neglecting total-derivative (boundary) terms~\cite{peskin2018introduction},
the action becomes
\begin{align}
S=&\int{dx}\Big\{\Psi_{1}^{\dagger}(x)[G^{-1}-\Sigma_1]\Psi_{1}(x)\!-\!\frac{1}{2m}\Big[{\bm \nabla}\Big(\frac{2\theta\!+\!\phi}{4}\Big)\Big]^2\!+\!\frac{|\Delta|^2}{U}\Big\}+\int{dx}\Big\{\Psi_{2}^{\dagger}(x)
[G^{-1}-\Sigma_2]\Psi_{2}(x)\!-\!\frac{1}{2m}\Big[{\bm \nabla}\Big(\frac{2\theta\!-\!\phi}{4}\Big)\Big]^2\!+\!\frac{|\Delta|^2}{U}\Big\}\nonumber\\
&+\int{dtd{\bf q}}\frac{2|\mu_+(t,{\bf q})|^2}{V_+(q)}+\int{dtd{\bf q}}\frac{2|\mu_-(t,{\bf q})|^2}{V_-(q)}+S_{\rm JC},
\end{align}  
where the Nambu spinor $\Psi_{j}^{\dagger}=(\psi_{j\uparrow}^{\dagger},\psi_{j\uparrow})$; the bare inverse Nambu Green's function
\begin{equation}
G^{-1}=i\partial_t\!-\!\xi_{{\hat {\bf k}}}\tau_3\!-\!|\Delta|\tau_1-{\bf v_{\hat {\bf k}}}\cdot{\bf p}_s,
\end{equation}
with ${\bf p}_s={\bm \nabla}\theta/2$, and the self-energy corrections are 
\begin{align}
\Sigma_1=\frac{\partial_t\theta(x)}{2}\tau_3\!+\!\frac{\partial_t\phi(x)}{4}\tau_3\!+\!\mu_{+}\tau_3\!+\!\mu_-\tau_3\!+\!\frac{1}{2m}\Big[{\bm \nabla}\Big(\frac{2\theta\!+\!\phi}{4}\Big)\Big]^2\tau_3\!+\!\frac{{\bf v_{\hat {\bf k}}}\cdot{\bm \nabla}\phi}{4},\\
\Sigma_2=\frac{\partial_t\theta(x)}{2}\tau_3\!-\!\frac{\partial_t\phi(x)}{4}\tau_3\!+\!\mu_{+}\tau_3\!-\!\mu_-\tau_3\!+\!\frac{1}{2m}\Big[{\bm \nabla}\Big(\frac{2\theta\!-\!\phi}{4}\Big)\Big]^2\tau_3\!-\frac{{\bf v_{\hat {\bf k}}}\cdot{\bm \nabla}\phi}{4}.
\end{align}

The fermionic degrees of freedom can now be integrated out exactly,
yielding the effective action
\begin{align}
S_{\mathrm{eff}}=&\int{dx}\Big({{\rm \bar Tr}}\ln{G^{-1}}-\sum_n\frac{1}{n}{{\rm \bar Tr}}(\Sigma_1G)^n\Big)+\int{dx}\Big({{\rm \bar Tr}}\ln{G^{-1}}-\sum_n\frac{1}{n}{{\rm \bar Tr}}(\Sigma_2G)^n\Big)-\int{dx}\Big[\frac{1}{m}\Big(\frac{{\bm \nabla}\theta}{2}\Big)^2+\frac{1}{m}\Big(\frac{{\bm \nabla}\phi}{4}\Big)^2\Big]\nonumber\\
&+2\int{dx}\frac{|\Delta|^2}{U}+\int{dtd{\bf q}}\frac{2|\mu_+(t,{\bf q})|^2}{V_+(q)}+\int{dtd{\bf q}}\frac{2|\mu_-(t,{\bf q})|^2}{V_-(q)}+S_{\rm JC},
\end{align}
where the fermionic Green function in the Matsubara representation and momentum space is given by~\cite{abrikosov2012methods,mahan2013many,schrieffer1964theory} 
\begin{equation}\label{GreenfunctionMr}
G(p)=\frac{ip_n\tau_0-{\bf p}_s\cdot{\bf v_k}\tau_0+\xi_{\bf k}\tau_3+|\Delta|\tau_1}{(ip_n-E_{\bf k}^+)(ip_n-E_{\bf k}^-)}.  
\end{equation}  
Here, the quasiparticle spectrum acquires a Doppler shift  ${\bf v_{\bf k}}\cdot{\bf p}_{s}$~\cite{fulde1964superconductivity,yang2018fulde,yang21theory,PhysRevB.95.075304,PhysRevB.98.094507} and takes the form $E_{\bf k}^{\pm}={\bf v_{\bf k}}\cdot{{\bf p_{s}}}\pm{E_{\bf k}}$ where $E_{\bf k}=\sqrt{\xi_{\bf k}^2+|\Delta|^2}$ is the Bogoliubov dispersion.

Retaining the lowest two orders (i.e., $n=1$ and $n=2$) and neglecting total-derivative (boundary) terms yield
\begin{align}
S_{\mathrm{eff}}=&\,2\int{dR}\Big\{\sum_{p_n,{\bf k}}\ln[(ip_n-E_{\bf k}^+)(ip_n-E_{\bf k}^-)]+\frac{|\Delta|^2}{U}-\frac{\bar\chi_3}{2m}\Big(\frac{{\bm \nabla}\theta}{2}\Big)^2-\frac{\bar\chi_3}{2m}\Big(\frac{{\bm \nabla}\phi}{4}\Big)^2-\chi_{00}\frac{v_F^2}{2}\Big(\frac{{\bm\nabla}\phi}{4}\Big)^2\Big\}\nonumber\\
&-\int{dR}\chi_{33}\Big\{\Big[\frac{\partial_t\theta(x)}{2}\!+\!\frac{\partial_t\phi(x)}{4}\!+\!\mu_{+}\!+\!\mu_-\Big]^2+\Big[\frac{\partial_t\theta(x)}{2}\!-\!\frac{\partial_t\phi(x)}{4}\!+\!\mu_{+}\!-\!\mu_-\Big]^2\Big\}\nonumber\\
&+\int{dtd{\bf q}}\frac{2|\mu_+(t,{\bf q})|^2}{V_+(q)}+\int{dtd{\bf q}}\frac{2|\mu_-(t,{\bf q})|^2}{V_-(q)}+S_{\rm JC}\nonumber\\
=&\,2\int{dR}\Big\{\sum_{p_n,{\bf k}}\ln[(ip_n-E_{\bf k}^+)(ip_n-E_{\bf k}^-)]+\frac{|\Delta|^2}{U}-\frac{\bar\chi_3}{2m}\Big(\frac{{\bm \nabla}\theta}{2}\Big)^2-\frac{\bar\chi_3}{2m}\Big(\frac{{\bm \nabla}\phi}{4}\Big)^2-\chi_{00}\frac{v_F^2}{2}\Big(\frac{{\bm\nabla}\phi}{4}\Big)^2\Big\}\nonumber\\
&-2\int{dR}\chi_{33}\Big\{\Big[\frac{\partial_t\theta(x)}{2}\!+\!\mu_{+}\Big]^2+\Big[\!\frac{\partial_t\phi(x)}{4}\!+\!\mu_-\Big]^2\Big\}+\int{dtd{\bf q}}\frac{2|\mu_+(t,{\bf q})|^2}{V_+(q)}+\int{dtd{\bf q}}\frac{2|\mu_-(t,{\bf q})|^2}{V_-(q)}+S_{\rm JC},
\end{align}
with the correlation coefficients, 
\begin{eqnarray}
{\tilde \chi}_3&=&\sum_{\bf k}\Big[1+\sum_{p_n}{\rm Tr}[G_0(p)\tau_3]\Big]=\sum_{\bf k}\Big[1+\frac{\xi_{\bf k}}{E_{\bf k}}\Big(f(E_{\bf k}^+)-f(E_{\bf k}^-)\Big)\Big]=-\frac{k_F^2}{2m}\sum_{\bf k}\partial_{\xi_{\bf k}}\Big[\frac{\xi_{\bf k}}{E_{\bf k}}\Big(f(E_{\bf k}^+)-f(E_{\bf k}^-)\Big)\Big]=n,\quad\quad \\
\chi_{03}&=&\frac{1}{2}\sum_p{\rm Tr}[G_0(p)\tau_0G_0(p)\tau_3]=0,\\
\chi_{33}&=&\frac{1}{2}\sum_p{\rm Tr}[G_0(p)\tau_3G_0(p)\tau_3]=\sum_{p_n,{\bf k}}\frac{(ip_n-{\bf p}_s\cdot{\bf v_k})^2+\xi_{\bf k}^2-|\Delta|^2}{(ip_n-E_{\bf k}^+)^2(ip_n-E_{\bf k}^-)^2}=\sum_{\bf k}\partial_{\xi_{\bf k}}\Big[\frac{\xi_{\bf k}}{2E_{\bf k}}\Big(f(E_{\bf k}^+)-f(E_{\bf k}^-)\Big)\Big]=-D,\\
\chi_{00}&=&\frac{1}{2}\sum_p{\rm Tr}[G_0(p)G_0(p)]=\sum_{p_n,{\bf k}}\frac{(ip_n-{\bf p}_s\cdot{\bf v_k})^2+\xi_{\bf k}^2+|\Delta|^2}{(ip_n-E_{\bf k}^+)^2(ip_n-E_{\bf k}^-)^2}=\sum_{\bf k}\partial_{E_{\bf k}}\Big[\frac{f(E_{\bf k}^+)-f(E_{\bf k}^-)}{2}\Big].~~~~~
\end{eqnarray}
Here, $n$ is carrier density and $D$ is the density of states of carriers; $R=(t,{\bf R})$ denotes the center-of-mass coordinate~\cite{schrieffer1964theory,yang21theory}, where $\mathbf{R}$ is confined to the 2D in-plane space of each layer.

Integrating out the Hartree fields yields the effective action for the
gap and phase fluctuations:
\begin{align}
S_{\rm eff}=&\,2\int{dR}\Big\{\sum_{p_n,{\bf k}}\ln[(ip_n-E_{\bf k}^+)(ip_n-E_{\bf k}^-)]+\frac{|\Delta|^2}{U}-\frac{np_s^2}{2m}\Big\}+2\int{dt}d{\bf q}\Big\{\frac{D}{1+DV_{+}(q)}\Big[\frac{\partial_t\theta(t,{\bf q})}{2}\Big]^2\Big\}\nonumber\\
&+2\int{dt}d{\bf q}\Big\{\frac{D}{1+DV_{-}(q)}\Big[\!\frac{\partial_t\phi(t,{\bf q})}{4}\Big]^2\Big\}-2\int{dR}\frac{f_s}{2}\Big(\frac{{\bm \nabla}\phi}{4}\Big)^2+S_{\rm JC},
\end{align}
where the superfluid stiffness can be expressed as
\begin{eqnarray}
f_s&=&\frac{\chi_3}{m}+v_F^2\chi_{00}=v_F^2\sum_{\bf k}\Big\{\partial_{E_{\bf k}}\Big[\frac{f(E_{\bf k}^+)-f(E_{\bf k}^-)}{2}\Big]-\partial_{\xi_{\bf k}}\Big[\frac{\xi_{\bf k}}{2E_{\bf k}}\Big(f(E_{\bf k}^+)-f(E_{\bf k}^-)\Big)\Big]\Big\}=v^2_F\sum_{\bf k}\frac{|\Delta|^2}{E_{\bf k}}\Big[\frac{f(E_{\bf k}^+)-f(E_{\bf k}^-)}{2E_{\bf k}}\Big]\nonumber\\
&\approx&v^2_F{|\Delta|^2}\sum_{\bf k}\frac{f(E_{\bf k}^-)-f(E_{\bf k}^+)}{2E^3_{\bf k}}.
\end{eqnarray}
In the long-wavelength limit, one has
\begin{equation}
V_{+}(q)=\frac{2\pi e^2}{\varepsilon q}(1+e^{-qd})\approx \frac{4\pi e^2}{\varepsilon q}=2V_{\rm 2D}(q), \quad\quad V_{-}(q)=\frac{2\pi e^2}{\varepsilon q}(1-e^{-qd})\approx \frac{2\pi e^2d}{\varepsilon},
\end{equation}
and then, the effective action is simplified to 
\begin{equation}
S_{\rm eff}= S_{\rm intra}\big(|\Delta|,{\bf p}_s^{2}\big) + S_J(\phi).
\end{equation} 
where the intralayer contribution reads 
\begin{equation}
S_{\rm intra}\big(|\Delta|,{\bf p}_s^{2}\big)=2\int{dR}\Big\{\sum_{p_n,{\bf k}}\ln[(ip_n-E_{\bf k}^+)(ip_n-E_{\bf k}^-)]+\frac{|\Delta|^2}{U}-\frac{np_s^2}{2m}\Big\}+2\int{dt}d{\bf q}\Big\{\frac{D}{1+2DV_{\rm 2D}(q)}\Big[\frac{\partial_t\theta(t,{\bf q})}{2}\Big]^2\Big\}
\end{equation}
while the Josephson contribution is given by
\begin{equation}\label{JPac}
S_J(\phi)=\int{dR}\Big[\frac{D_l}{2}\Big(\frac{\partial_t\phi}{2}\Big)^2-\frac{f_s}{4}\Big(\frac{{\bm \nabla}\phi}{2}\Big)^2+|J_1||\Delta|^2\cos(\phi+\alpha)+|J_2||\Delta|^4\cos(2\phi)\Big],
\end{equation}
with $D_l=D/(1+D{2\pi e^2d}/{\varepsilon})$ being the effective density of states renormalized by the
interlayer Coulomb interaction.

The intralayer sector here coincides with the standard
effective action of a two-dimensional superconductor and
governs the intralayer superconducting properties~\cite{yang2025tractable,yang2025efficient,yang2025preformed,yang21theory}.
These contributions will be discussed in Sec.~\ref{secintralayer}.
In contrast, the Josephson sector $S_J(\phi)$ controls the
equilibrium configuration and dynamical behavior of the
relative phase between the two layers,
which will be analyzed in Sec.~\ref{secjf}. The resulting superfluid stiffness derived here  is fully consistent with those obtained from Gor'kov theory, kinetic formulations, and current–current correlation approaches~\cite{abrikosov2012methods,yang21theory,yang19gauge,PhysRevB.98.094507,sun20collective,PhysRevB.106.144509}.

\subsubsection{Inclusion of disorder}

The above derivation holds in the clean limit. In practice, disorder is
unavoidable in monolayer systems and affects the gap and phase (stiffness) sectors in qualitatively different ways. For isotropic $s$-wave pairing in the diffusive limit, nonmagnetic disorder does not directly modify the gap equation, in accordance with Anderson’s theorem. Instead, disorder influences the superconducting state primarily through the renormalization of the superfluid phase stiffness. This conclusion has been consistently established using multiple theoretical frameworks, including Gor'kov formalism~\cite{abrikosov2012methods}, Eilenberger transport theory~\cite{eilenberger1968transformation,PhysRevLett.25.507,PhysRevB.99.224511}, gauge-invariant kinetic equations~\cite{PhysRevB.98.094507}, and diagrammatic approaches incorporating vertex corrections in the current–current response~\cite{PhysRevB.99.224511,PhysRevB.106.144509}. The essential reason is that the superfluid stiffness is determined by the transverse current–current response kernel. In the presence of impurity scattering, vertex corrections necessarily enter this kernel and lead to a suppression of the stiffness by elastic scattering processes.

To incorporate this effect without explicitly solving the fully impurity-dressed response functions, one may adopt the well-established clean-to-dirty interpolation introduced by Tinkham~\cite{tinkham2004introduction}. This interpolation relates the penetration depth via $\lambda^2=\lambda_{\rm clean}^2(1+\xi/l)$, which implies $n_s\rightarrow n_{s,{\rm clean}}/(1+\xi/l)$. This approach  recovers the Mattis–Bardeen dirty-limit behavior~\cite{PhysRev.111.412,PhysRevB.109.064508}, connects smoothly to finite-scattering extensions~\cite{PhysRev.156.470}, quantitatively captures penetration-depth measurements across disorder regimes in conventional superconductors~\cite{PhysRevB.76.094515,PhysRevB.72.064503,PhysRevB.10.1885,PhysRevB.6.2579}, and is further supported by fully microscopic derivations. Then, the superfluid stiffness takes the form  
 \begin{equation}\label{SD2}
f_s=\frac{v_F^2|\Delta|^2}{1+\xi/l}\sum_{\bf k}\frac{f(E_{\bf k}^-)-f(E_{\bf k}^+)}{2E^3_{\bf k}},
 \end{equation}
where $\xi=\hbar v_F/|\Delta|$ is the coherence length and $l=v_F\tau$ is the mean free path  with $\tau$ being the effective scattering time.

\section{Derivation of Josephson fluctuations}
\label{secjf}

In this section, we derive the effective Josephson dynamics in the presence of phase fluctuations from a microscopic perspective. Specifically, we decompose the Josephson phase as
 \begin{equation}\label{dec}
 \phi=\phi_e+\delta\phi(R),
 \end{equation}
 where $\phi_e$ denotes the equilibrium phase configuration, and $\delta\phi(R)$ describes dynamical fluctuations around this equilibrium. Substituting this decomposition into the Josephson-phase action enables a systematic and self-consistent separation between the static equilibrium configuration and the dynamical fluctuations. Most importantly, this procedure allows us to bosonize the Josephson phase fluctuations and treat them as collective bosonic modes, thereby capturing fluctuation-induced renormalizations of the Josephson response in a controlled, fully microscopic framework that goes beyond both mean-field approximations and phenomenological descriptions.

\subsection{Equilibrium configuration}

Substituting the decomposition [Eq.~(\ref{dec})] into the Josephson-phase-related action $S_{J}(\phi)$ [Eq.~(\ref{JPac})], the optimal equilibrium phase $\phi^{\rm op}_e$ is determined by minimizing the static Josephson energy obtained from the potential part of the action, 
\begin{eqnarray}
E_J(\phi_e)
&=&
-|J_1||\Delta|^2
\big\langle \cos(\phi+\alpha) \big\rangle
-|J_2||\Delta|^4
\langle\cos(2\phi)\rangle.
\end{eqnarray}
Using the statistic average $\langle\delta\phi\rangle=0$ and $\langle\delta\phi^2\rangle\ne0$, the thermal average can be evaluated via the cumulant expansion~\cite{mahan2013many,abrikosov2012methods,peskin2018introduction}, 
\begin{align}&\langle\cos(m\phi)\rangle={\rm Re}[e^{im\phi_e}\langle{e^{im\delta\phi}}\rangle]={\rm Re}\Big[e^{im\phi_e}\sum_{n=0}\frac{\langle(im\delta\phi)^{2n}\rangle}{(2n)!}\Big]={\rm Re}\Big[e^{im\phi_e}\sum_{n=0}\frac{(2n-1)!!(-1)^nm^{2n}\langle\delta\phi^{2}\rangle^n}{(2n)!}\Big]\nonumber\\
&={\rm Re}\Big[e^{im\phi_e}\sum_{n=0}\frac{(-1)^nm^{2n}\langle\delta\phi^{2}\rangle^n}{2^nn!}\Big]={\rm Re}\Big[e^{im\phi_e}e^{-\langle{\delta\phi}^2\rangle/2}\Big]=\cos(m\phi_e)e^{-m^2\langle{\delta\phi}^2\rangle/2}.
\end{align}
As a result, the fluctuation-renormalized Josephson energy becomes
\begin{eqnarray}
E_J(\phi_e)
&=&
-{\bar J}_1(T)\cos(\phi_e+\alpha)-
{\bar J}_2(T)
\cos(2\phi_e),\label{SIEJ}
\end{eqnarray}
where the effective Josephson couplings are renormalized by Josephson phase fluctuations as 
\begin{eqnarray}
{\bar J}_1(T)&=&|J_1||\Delta(T)|^2\exp[-\langle \delta\phi^2(T)\rangle/2],\label{SIJ1} \\
{\bar J}_2(T)&=&|J_2||\Delta(T)|^4\exp[-2\langle \delta\phi^2(T)\rangle]. \label{SIJ2}
\end{eqnarray}
Consequently, as rigorously demonstrated above, 
the optimal equilibrium phase $\phi^{\rm op}_e$ 
 is determined by minimizing the   fluctuation-renormalized Josephson energy rather than the bare one,
 \begin{equation}\label{SIMi}
 \phi_e^{\rm op}=\arg\min_{\phi^e\in[0,2\pi)}E_J(\phi_e),
 \end{equation}
 suggesting the nontrivial impact of phase fluctuations on the equilibrium configuration.

\subsubsection{Josephson current}

In the presence of an externally imposed dc Josephson current $I_J$, the thermodynamic potential acquires an additional work term associated with the phase bias. The effective Josephson energy is therefore modified to~\cite{mahan2013many}
\begin{eqnarray}
\tilde{E}_J(\phi_e)
&=&
-{\bar J}_1(T)\cos(\phi_e+\alpha)-
{\bar J}_2(T)
\cos(2\phi_e)-\frac{I_J}{2e}\phi_e,
\end{eqnarray}
The equilibrium phase under a finite current is determined by minimizing $\tilde E_J(\phi_e)$. This yields the generalized current–phase relation 
\begin{equation}\label{SIcu}
I_J(\phi_e)=2e[{\bar J}_1(T)\sin(\phi_e+\alpha)+2{\bar J}_2(T)\sin(2\phi_e)].
\end{equation}
The phase fluctuations thus renormalize not only the Josephson energy but also the current–phase relation, suppressing both harmonic components exponentially via the Debye–Waller-like factors~\cite{kittel1963quantum,sakurai2020modern}.

Particularly, higher-order Josephson couplings are suppressed more strongly than the lowest-order term: the second-harmonic coupling  ${\bar J}_2\propto\exp({-2\langle{\delta\phi^2}\rangle})$  decays parametrically faster than the first-harmonic coupling ${\bar J}_1\propto\exp({-\langle{\delta\phi^2}\rangle}/2)$  as phase fluctuations increase with temperature. Thus, non-sinusoidal components of the Josephson current–phase relation are particularly fragile against phase decoherence, with higher-order harmonics expected to be even more fragile.  As the Josephson diode nonreciprocity heavily  relies on these non-sinusoidal components~\cite{Pal2022,PhysRevB.109.094518,Can2021}, the Josephson diode effect is expected to disappear at a temperature lower than the critical temperature where the Josephson critical current is suppressed.

\subsection{Dynamics of Josephson phase fluctuations}

Then, we derive the dynamics of the Josephson phase fluctuations. 
Starting from Eq.~(\ref{JPac}) and the decomposition in Eq.~(\ref{dec}), the
Euler–Lagrange equation with respect to the phase fluctuation field $\delta\phi$ reads~\cite{peskin2018introduction}
\begin{equation}
\partial_{\mu}\left[\frac{\partial S_J[\phi]}{\partial(\partial_{\mu}\delta\phi/2)}\right]
=\frac{\partial S_J[\phi]}{\partial(\delta\phi/2)} ,
\end{equation}
which leads to the equation of motion
\begin{equation}
[D_l\partial_t^2-f_s{\bf \nabla}^2/2]\delta\phi/2+2|J_1||\Delta|^2\sin(\phi_e+\delta\phi+\alpha)+4|J_2||\Delta|^4\sin(2\phi_e+2\delta\phi)=0.
\end{equation}

To proceed, we treat the phase fluctuations within a field-theoretical cumulant expansion~\cite{peskin2018introduction,abrikosov2012methods},
retaining only a single phase operator while contracting the remaining fields.
Using the trigonometric identity
\begin{align}
\sin(m\phi_e+m\delta\phi)
=\sin{m}\phi_e\cos{m}\delta\phi+\cos{m}\phi_e\sin{m}\delta\phi ,
\end{align}
and assuming $\langle\delta\phi\rangle=0$, the critical response in the equation of motion is governed by
the second contribution. The sine of the phase fluctuation is expanded as
\begin{align}
\sin{m}\delta\phi(R)
=\sum_{n=0}^{\infty}\frac{(-1)^n{m}^{2n+1}}{(2n+1)!}\,\delta\phi^{\,2n+1}(R).
\end{align}
Within the cumulant approximation~\cite{peskin2018introduction,mahan2013many,abrikosov2012methods}, we retain only a single uncontracted phase operator
while contracting the remaining $2n$ fields according to Wick's theorem~\cite{peskin2018introduction},
\begin{align}
\delta\phi^{\,2n+1}(R)\;\rightarrow\;(2n-1)!!\,\langle\delta\phi^2(R)\rangle^{n}\,\delta\phi(R).
\end{align}
Substituting this back, we obtain
\begin{align}
\sin\delta\phi(R)
&\rightarrow \sum_{n=0}^{\infty}\frac{(-1)^n{m}^{2n+1}}{(2n+1)!}\,(2n-1)!!\,\langle\delta\phi^2(R)\rangle^{n}\,\delta\phi(R)\nonumber\\
&=\sum_{n=0}^{\infty}\frac{(-1)^nm^{2n+1}}{2^n n!}\,\langle\delta\phi^2(R)\rangle^{n}\,\delta\phi(R)
=\exp\!\Big[-\frac{m^2}{2}\langle\delta\phi^2\rangle\Big]m\delta\phi(R).
\end{align}
Then, the equation of motion of the Josephson phase fluctuations reads
\begin{equation}
\Big[D_l\partial_t^2-{f_s}{\bf \nabla}^2/2+4|J_1||\Delta|^2\exp(-\langle\delta\phi^2\rangle/2)\cos(\phi_e+\alpha)+16|J_2||\Delta|^4\exp(-2\langle\delta\phi^2\rangle)\cos(2\phi_e)\Big]\frac{\delta\phi(R)}{2}=0,
\end{equation}
or equivalently, 
\begin{equation}\label{EOM}
\Big[D_l\partial_t^2-{f_s}{\bf \nabla}^2/2+4{\bar J}_1\cos(\phi_e+\alpha)+16{\bar J}_2\cos(2\phi_e)\Big]\frac{\delta\phi}{2}=0.
\end{equation}
From this equation, the energy spectrum of the collective Josephson phase mode follows as
\begin{equation}\label{SIom}
D_l\omega^2_{\rm L}({q})={4{\bar J}_1\cos(\phi^{\rm op}_e\!+\!\alpha)}+{16{\bar J}_2\cos(2\phi^{\rm op}_e)}
+{f_sq^2}/{2}.
\end{equation}
Clearly, the interlayer Josephson coupling opens a finite gap in the collective excitation spectrum by locking the relative phase, in agreement with the Leggett-type phase mode~\cite{Leggett1966,sun20collective}. 

In particular, the renormalized Josephson couplings ${\bar J}_1(T)$ and ${\bar J}_2(T)$ entering the excitation gap explicitly encode the feedback between thermal phase fluctuations and Josephson coherence. As temperature increases, the growth of the thermal phase fluctuation $\langle\delta\phi^2(T)\rangle$ leads to an exponential suppression of ${\bar J}_{1,2}(T)$ through the Debye–Waller-like factor, while the reduction of the pairing amplitude $|\Delta(T)|$ further suppresses the couplings via their explicit amplitude dependence. The resulting decrease in Josephson stiffness produces a progressive softening of the phase mode upon warming. The softened mode in turn enhances Josephson phase fluctuations, establishing a self-consistent feedback mechanism.

\subsection{Statistic average of Josephson phase fluctuations}

From the equation of motion of the Josephson phase fluctuations in Eq.~(\ref{EOM}),
the average of Josephson phase fluctuations can be obtained within the quantum-statistic mechanism~\cite{abrikosov2012methods,mahan2013many} either via the fluctuation--dissipation theorem
or within the Matsubara formalism.
The two approaches are fully equivalent and yield identical results,
\begin{equation}\label{SIe}
\langle \delta\phi^2(T)\rangle=\int\frac{2d{\bf q}}{(2\pi)^2}\frac{2n_B[\omega_{\rm L}({q})]+1}{D_l\omega_{\rm L}({q})},
\end{equation}
reflecting the bosonic description of the phase degree of freedom as a collective phase mode.\\

{\sl Fluctuation dissipation theorem.---}Considering the thermal fluctuations, the dynamics of the phase from Eq.~(\ref{EOM}) is given by
\begin{equation}
\big[D_l\omega_L^2({\bf q})-\omega^2+iD_l\omega{\gamma}\big]\frac{\delta\phi({\omega,{\bf q}})}{2}={J}_{\rm th}(\omega,{\bf q}).    
\end{equation}  
Here, we have introduced a thermal field ${J}_{\rm th}(\omega,{\bf q})$ that obeys the fluctuation-dissipation theorem~\cite{Landaubook}:
\begin{equation}
\langle{J}_{\rm th}(\omega,{\bf q}){J}^*_{\rm th}(\omega',{\bf q}')\rangle=\frac{(2\pi)^3D_l\gamma\omega\delta(\omega-\omega')\delta({\bf q-q'})}{\tanh(\beta\omega/2)},   
\end{equation}
and $\gamma=0^+$ is a phenomenological damping constant. From this dynamics, the average of the phase fluctuations is given by
\begin{eqnarray}
  \frac{1}{4}\langle{\delta\phi^2}\rangle&=&\int\frac{d\omega{d\omega'}d{\bf q}d{\bf q'}}{(2\pi)^6}\frac{\langle{J}_{\rm th}(\omega,{\bf q}){J}^*_{\rm th}(\omega',{\bf q}')\rangle}{D^2_l\big[\big(\omega^2\!-\!\omega_L^2({\bf q})\big)\!-\!i\omega\gamma\big]\big[({\omega'}^2\!-\!\omega_L^2({\bf q}')\big)\!+\!i\omega'\gamma\big]}=\int\frac{d\omega{d{\bf q}}}{D_l(2\pi)^2}\frac{\gamma\omega/\tanh(\beta\omega/2)}{\big[\omega^2\!-\!\omega_L^2({\bf q})\big]^2+\omega^2\gamma^2}\nonumber\\
&=&\int\frac{{d{\bf q}}}{(2\pi)^2}\frac{[2n_B\big(\omega_L({\bf q})\big)+1]}{2D_l\omega_L({\bf q})}.\label{EEEE}
\end{eqnarray}

{\sl Matsubara formalism.---}Within the Matsubara formalism, by mapping into the imaginary-time space, from Eq.~(\ref{EOM}),  the thermal phase fluctuation reads~\cite{abrikosov2012methods}
\begin{eqnarray}
   \frac{1}{4}\langle{\delta\phi^2}\rangle&\!=\!&\int\frac{d{\bf q}}{(2\pi)^2}\Big[\Big\langle\Big|\frac{\delta\phi^*(\tau,{\bf q})}{2}\frac{\delta\phi(\tau,{\bf q})}{2}e^{-\int^{\beta}_{0}d\tau{d{\bf q}}D_l{\delta\phi^*(\tau,{\bf q})}(\omega_L^2({\bf q})-\partial_{\tau}^2){\delta\phi(\tau,{\bf q})}/4}\Big|\Big\rangle\Big]\nonumber\\
  &\!=\!&\int\frac{d{\bf q}}{(2\pi)^2}\Big[\frac{1}{\mathcal{Z}_0}{\int}D\delta\phi{D\delta\phi^*}\frac{\delta\phi^*(\tau,{\bf q})}{2}\frac{\delta\phi(\tau,{\bf q})}{2}e^{-\int^{\beta}_{0}d\tau{d{\bf q}}D_l{\delta\phi^*(\tau,{\bf q})}(\omega_L^2({\bf q})-\partial_{\tau}^2){\delta\phi(\tau,{\bf q})}/4}\Big]\nonumber\\
   &=&\!\!\!\int\frac{d{\bf q}}{(2\pi)^2}\frac{1}{\mathcal{Z}_0}{\int}D\delta\phi{D\delta\phi^*}\delta_{J^*_{\bf q}}\delta_{J_{\bf q}}e^{-\int^{\beta}_{0}d\tau{d{\bf q}}[D_l{\delta\phi^*(\tau,{\bf q})}(\omega_L^2({\bf q})-\partial_{\tau}^2){\delta\phi(\tau,{\bf q})}/4+J_{\bf q}\delta\phi({\bf q})/2+J^*_{\bf q}\delta\phi^*({\bf q})/2]}\Big|_{J=J^*=0}\nonumber\\
  &=&\!\!\!\int\frac{d{\bf q}}{(2\pi)^2}\frac{1}{D_l}\delta_{J^*_{\bf q}}\delta_{J_{\bf q}}\exp\Big\{-\int_0^{\beta}{d\tau}\sum_{q'}J_{q'}\frac{1}{\partial_{\tau}^2-\omega_L^2({\bf q})}J^*_{q'}\Big\}\Big|_{J=J^*=0}=-\int\frac{d{\bf q}}{(2\pi)^2}\frac{1}{\beta}\sum_{\omega_n}\frac{1}{D_l}\frac{1}{(i\Omega_n)^2\!-\!\omega_L^2({\bf q})}\nonumber\\
 &=&\int\frac{{d{\bf q}}}{(2\pi)^2}\frac{2n_B(\omega_L({\bf q}))+1}{2D_l\omega_L({\bf q})},\label{mf}
\end{eqnarray}
which is exactly the same as the one in Eq.~(\ref{EEEE}) obtained via the fluctuation dissipation theorem. Here, $\Omega_n=2n\pi{T}$ represents the Bosonic Matsubara frequencies; ${J_{\bf q}}$ denotes the generating functional and $\delta{J_{\bf q}}$ stands for the functional derivative.\\
\section{Intralayer superconducting properties: framework beyond mean-field theory}
\label{secintralayer}

The effective intralayer action $S_{\rm intra}$ governs the superconducting (SC) properties within each individual layer. 
From this action, the temperature-dependent gap amplitude $|\Delta(T)|$, the phase-fluctuation correlator $\langle {\bf p}_s^{2} \rangle$, and the renormalized superfluid stiffness $f_s(T)$ is determined at the fully microscopic level.  Within our gauge-invariant formulation~\cite{benfatto10,yang19gauge}, the phase fluctuations encoded in $\langle {\bf p}_s^{2} \rangle$ correspond to the in-phase collective excitation~\cite{sun20collective,yang19gauge,yang21theory}, namely the bosonic Nambu–Goldstone (NG) mode associated with spontaneous U(1) symmetry breaking in the superconducting state~\cite{nambu1960quasi,goldstone1961field,goldstone1962broken,nambu2009nobel}. 
These fluctuations are coupled with long-range Coulomb interactions in the symmetric channel, as demonstrated in Sec.~\ref{seceac}.

In 3D superconductors, the NG mode is rendered dynamically inert due to the Anderson–Higgs mechanism~\cite{anderson63plasmons,ambegaokar61electromagnetic,littlewood81gauge}, acquiring a plasma gap and therefore not contributing to low-energy thermodynamics. 
In contrast, in 2D systems, the mode remains gapless and constitutes an active low-energy bosonic excitation~\cite{yang21theory,PhysRevB.97.054510,yang2025tractable,yang2025efficient,yang2025preformed}. 
As a result, phase fluctuations play a central role in determining the thermodynamic and spectral properties of 2D superconductors. In our recent work on 2D superconductors~\cite{yang2025tractable,yang2025efficient,yang2025preformed}, we developed a fully self-consistent framework that incorporates long-wavelength NG-mode fluctuations into the superconducting gap equation in the presence of long-range Coulomb interactions, going beyond the conventional mean-field description. 
A key feature of this approach is that, although the average superfluid momentum vanishes,
\begin{equation}
\langle {\bf p}_s \rangle = 0,
\end{equation}
the fluctuation correlator remains finite,
\begin{equation}
\langle {\bf p}_s^{2} \rangle \neq 0.
\end{equation}
These finite fluctuations are carried by \emph{bosonic} NG excitations and enter the fermionic self-energy and gap equation in a self-consistent manner.  Consequently, both the superconducting gap $|\Delta(T)|$ and its disappearance temperature $T_s$ acquire nontrivial dependence on the strength of the disorder and the carrier density, reflecting fluctuations-induced renormalization effects beyond the mean-field theory~\cite{yang2025tractable,yang2025efficient,yang2025preformed}.

In the present work, we directly adopt this fluctuation-renormalized framework for the in-plane sector.  By solving the coupled gap and NG-phase-sector equations self-consistently, we determine the superconducting gap $|\Delta(T)|$ and the renormalized phase stiffness $f_s(T)$. These quantities then serve as the fundamental inputs for the subsequent analysis of interlayer Josephson coupling.

Specifically, the self-consistent gap equation takes the form~\cite{yang2025tractable,yang2025efficient,yang2025preformed,yang21theory}
\begin{equation}\label{OGE}
\frac{1}{U}=F({p}_s^{2},|\Delta|,T)
=\sum_{\bf k}\frac{f(E_{\bf k}^+)-f(E_{\bf k}^-)}{2E_{\bf k}},
\end{equation}
where $f(x)$ is the Fermi distribution function. 
The quasiparticle spectrum acquires a Doppler shift ${\bf v}_{\bf k}\!\cdot\!{\bf p}_s$~\cite{fulde1964superconductivity,yang2018fulde,yang21theory,PhysRevB.95.075304,PhysRevB.98.094507} and takes the form
\begin{equation}
E_{\bf k}^{\pm}={\bf v}_{\bf k}\!\cdot\!{\bf p}_s \pm E_{\bf k},
\end{equation}
with $E_{\bf k}=\sqrt{\xi_{\bf k}^2+|\Delta|^2}$ the Bogoliubov dispersion, 
$\xi_{\bf k}=\hbar^2k^2/(2m)-\mu$, 
and ${\bf v}_{\bf k}=\partial_{\bf k}\xi_{\bf k}$ the band velocity. In this gauge-invariant formulation, NG phase fluctuations enter the gap equation through the Doppler term in a manner formally analogous to an electromagnetic vector potential~\cite{ambegaokar61electromagnetic,nambu1960quasi,nambu2009nobel}. 
Since the Doppler shift appears linearly but the gap equation involves the symmetric combination $f(E_{\bf k}^+)-f(E_{\bf k}^-)$, the leading contribution of phase fluctuations is controlled by the second moment $\langle p_s^2\rangle$. 

The NG fluctuations contribute through the mean-squared superfluid momentum~\cite{yang2025tractable,yang2025efficient,yang2025preformed,yang21theory},
\begin{equation}\label{OPEF}
\langle p_s^2\rangle
=\int\frac{d{\bf q}}{(2\pi)^2}
\frac{q^2\,[2n_B(\omega_{\rm NG})+1]}
{D_q\,\omega_{\rm NG}(q)},
\end{equation}
where $n_B(x)$ is the Bose–Einstein distribution. 
The NG spectrum is given by~\cite{yang2025tractable,yang2025efficient,yang2025preformed,yang21theory}
\begin{equation}
\omega_{\rm NG}(q)=\sqrt{\frac{f_s q^2}{2D_q}},
\end{equation}
with $D_q=D/[1+2DV_{\rm 2D}(q)]$ the Coulomb-renormalized phase stiffness kernel. In the long-wavelength limit, the NG spectrum $\omega_{\rm NG}(q)\propto \sqrt{q}$, consistent with various  theoretical studies of 2D superconductors~\cite{sun20collective,yang21theory,ambegaokar61electromagnetic,PhysRevB.64.140506,PhysRevB.69.184510,PhysRevB.70.214531,PhysRevB.97.054510}.

The superfluid stiffness $f_s$, obtained from the NG-phase sector, coincides with that derived from the phase-twist (Josephson) response, reflecting the internal consistency of the gauge-invariant formulation. It is expressed as
\begin{equation}\label{OSD}
f_s=
\frac{v_F^2|\Delta|^2}{1+\xi/l}
\sum_{\bf k}
\frac{f(E_{\bf k}^-)-f(E_{\bf k}^+)}{2E_{\bf k}^3},
\end{equation}
where $\xi=\hbar v_F/|\Delta|$ is the superconducting coherence length and $l=v_F\tau$ the mean free path ($\tau$ is the effective scattering time). 
The prefactor $(1+\xi/l)^{-1}$ accounts for disorder-induced renormalization of the phase stiffness.

Consequently, Eqs.~(\ref{OGE})–(\ref{OSD}) form a closed set of coupled self-consistent equations for $\Delta(T)$ and $f_s(T)$. 
They incorporate fermionic quasiparticle excitations, NG phase fluctuations, and disorder effects on equal footing. The detailed derivation of this framework and its numerical implementation can be found in Refs.~\cite{yang2025tractable,yang2025efficient,yang2025preformed}.

One of the important consequences of this framework is that both the zero-temperature gap $|\Delta(0)|$ and the gap-closing temperature $T_s$ become strongly dependent on carrier density and disorder strength once phase fluctuations are incorporated self-consistently. This behavior goes beyond conventional mean-field expectations (Anderson theorem~\cite{anderson1959theory,suhl1959impurity,skalski1964properties,andersen2020generalized}), where these quantities are primarily controlled by the pairing interaction alone. In the present framework, long-wavelength NG fluctuations renormalize the quasiparticle spectrum and superfluid stiffness, which in turn feed back into the gap equation. As a result, $|\Delta(0)|$ and $T_s$ acquire nontrivial dependence on carrier density and impurity scattering. As demonstrated in our previous works~\cite{yang2025tractable,yang2025efficient,yang2025preformed}, such fluctuation-induced renormalization is consistent with experimental observations in gate-tuned two-dimensional superconductors, including monolayer WTe$_2$~\cite{Song2024,PhysRevResearch.7.013224,Sajadi2018,Fatemi2018}, bilayer MoS$_2$~\cite{Zheliuk2019}, and thin-film disordered InO$_x$~\cite{Charpentier2025}, where both the superconducting gap magnitude and transition scales exhibit pronounced density and disorder dependence.

\subsubsection{BKT physics}

It should be emphasized that Eqs.~(\ref{OGE})–(\ref{OSD}) provide the microscopic input, $|\Delta(T)|$ and $f_s(T)$, for calculating the interlayer Josephson properties and the corresponding out-of-plane transport behavior.

If one is instead interested in in-plane transport properties, the bare stiffness $f_s$ obtained above serves as the initial condition for the standard Berezinskii–Kosterlitz–Thouless (BKT) renormalization-group (RG) flow~\cite{benfatto10,PhysRevB.110.144518,PhysRevB.80.214506,PhysRevB.77.100506,PhysRevB.87.184505}:
\begin{equation}\label{OBKT}
\frac{dK}{dL}=-K^2g^2, 
\qquad
\frac{dg}{dL}=(2-K)g,
\end{equation}
with initial conditions
\begin{equation}
K(L=0)=\frac{\pi\hbar^2 f_s}{4k_BT},
\qquad
g(L=0)=2\pi e^{-c_0K(L=0)},
\end{equation}
where $c_0=2/\pi$ in the two-dimensional limit~\cite{yang2025tractable,yang2025efficient,yang2025preformed}. 
The fully renormalized superfluid density after vortex screening is
\begin{equation}
\bar f_s=\frac{4k_BT}{\pi\hbar^2}K(L=\infty).
\end{equation}

For critical behavior in the in-plane sector, the superconducting transition temperature $T^{\rm intra}_c$ is determined by the BKT universal jump condition of the renormalized stiffness $\bar f_s$, while the gap-closing temperature $T_s$ is defined by $\Delta(T_s)=0$. 
The in-plane transport properties associated with this BKT criticality have been discussed in detail in our previous works~\cite{yang2025tractable,yang2025efficient,yang2025preformed} and are not the focus of the present study.  

Interestingly, since the interlayer Josephson coherence is governed by a distinct, explicitly gapped phase mode, the corresponding transition temperature $T^{\rm inter}_c$ (denoted $T_c$ in the main text) associated with out-of-plane transport does not need to coincide with the in-plane BKT transition temperature $T^{\rm intra}_c$. 
This separation of phase-coherence scales provides a natural route to anisotropic superconducting transport behavior and warrants further investigation.

\section{Simulation treatments}

Our numerical analysis consists of two complementary steps which together establish a self-consistent connection between the intralayer superconducting properties and the interlayer Josephson response.

First, following the numerical implementation developed in our previous work, we self-consistently solve Eqs.~(\ref{OGE})–(\ref{OSD}) to obtain the temperature-dependent gap amplitude $|\Delta(T)|$ and superfluid stiffness $f_s(T)$. These quantities serve as the fundamental input for the subsequent interlayer Josephson calculation.

Using this fluctuation-renormalized input, we then self-consistently solve the coupled Josephson equations, Eqs.~(\ref{SIEJ})–(\ref{SIMi}) and Eqs.~(\ref{SIom})-(\ref{SIe}), which determine the optimal equilibrium phase configuration and the thermal average of Josephson phase fluctuations.  Once the self-consistent solution is obtained, the current–phase relation is evaluated from Eq.~(\ref{SIcu}).

The specific parameter values employed in the simulations are summarized in Table~\ref{Table}. 
With the chosen parameters and in the absence of phase fluctuations (mean-field limit), the zero-temperature superconducting gap is $|\Delta_{\rm MF}(0)| = 2~\text{meV}$, and the Josephson-coupling-induced excitation gap of the interlayer collective mode equals $0.21\,|\Delta_{\rm MF}(0)|$, ensuring that the Josephson collective mode remains well-defined inside the superconducting gap.

\begin{table}[htb]
\centering
 \caption{
The pairing interaction strength $U$ is chosen such that the BCS constant $\lambda=0.55$. 
The Debye cutoff energy in the gap equation is taken to be $\omega_D = 6$~meV. 
The detailed numerical implementation for the intralayer superconducting properties can be found in Refs.~\cite{yang2025tractable,yang2025efficient,yang2025preformed}. 
The effective mass is set equal to the free electron mass, $m = m_e$.  For simplicity, we take the high-energy cutoff for the zero-point contribution of the Josephson bosonic excitations to be the same Debye scale, $\omega_D$. 
Owing to the Debye–Waller–like exponential structure of the phase-fluctuation renormalization, the zero-point contribution can, in fact, be absorbed into the zero-temperature Josephson parameters $\bar J_1(0)$ and $\bar J_2(0)$. 
As a result, the precise value of this ultraviolet cutoff does not affect the temperature dependence or qualitative conclusions of the present work. The electronic mean free path depends strongly on sample quality and disorder, as monolayer materials are particularly susceptible to disorder due to their reduced dimensionality. In our estimates, we consider a reasonable range 
$l\sim1$-300~nm.}
\renewcommand\arraystretch{1.5}   \label{Table}
\begin{tabular}{l l l l l l l l}
    \hline
    \hline
   $~~\lambda$&\quad\quad~$\omega_D$&\quad\quad\quad\quad~$J_1$&\quad\quad\quad\quad~$J_2$&\quad\quad\quad~$~\alpha$&\quad\quad\quad\quad\quad~$v_F$&\quad\quad\quad\quad~$~l$\\   
0.55&\quad\quad$6~$meV&\quad\quad~$16~$meV$^{-1}\mu$m$^{-2}$&\quad\quad~$1~$meV$^{-3}\mu$m$^{-2}$&\quad\quad\quad~$0.2\pi$&\quad\quad\quad~$[0.3,\,2]\times10^5~{\rm m/s}$&\quad\quad\quad~$[1,300]~$nm\\
    \hline
    \hline
\end{tabular}\\
\end{table}

\section{Zero-point oscillations}

\begin{figure}[H]
\centering
  \includegraphics[width=8.6cm]{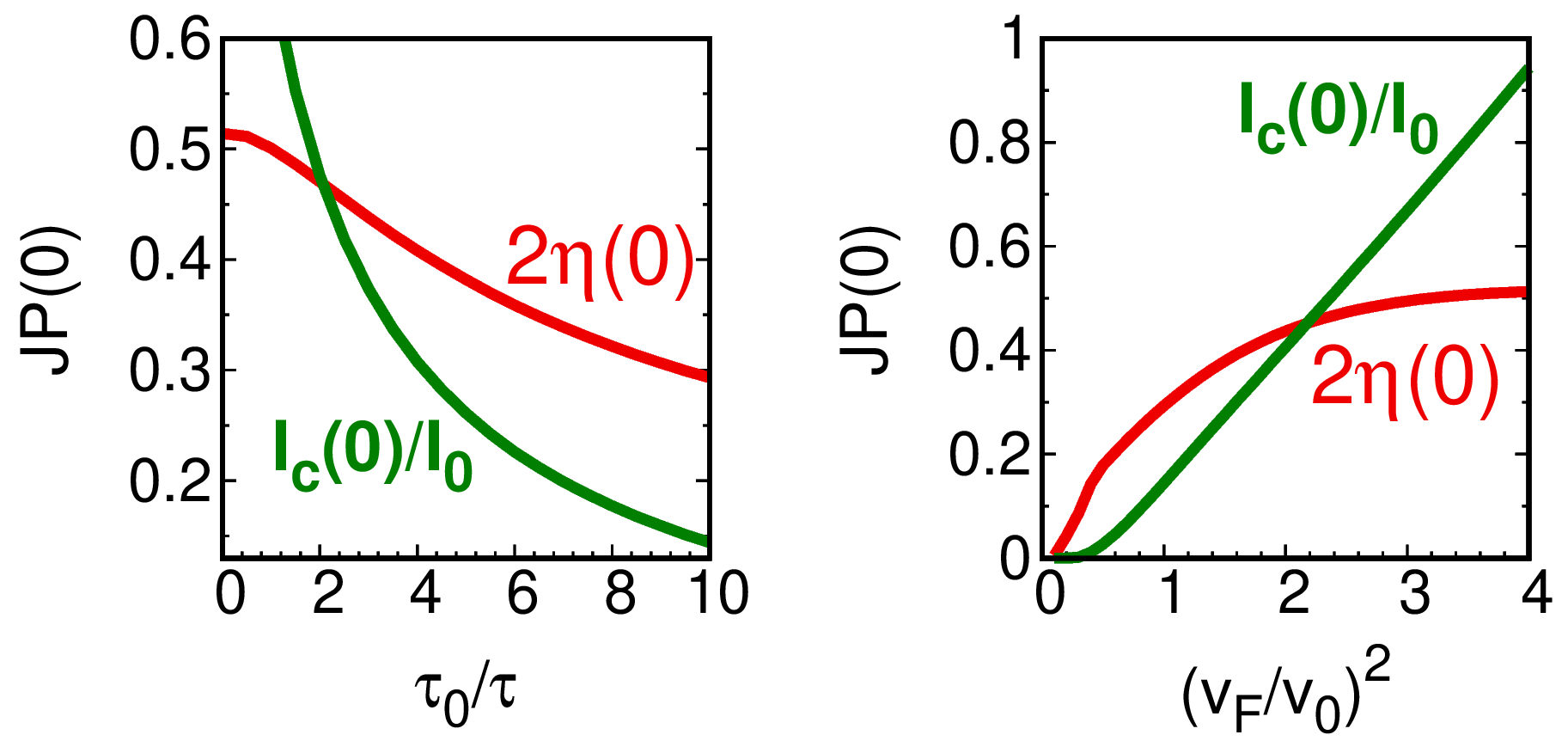}
\caption{
Zero-temperature Josephson critical current $I_c(0)$ and diode efficiency $\eta(0)$ under varying phase stiffness conditions. 
Left panel: Increasing disorder at $v_F=v_0$. 
Right panel: Reducing the Fermi velocity at $\tau=0.1\tau_0$. Here $v_0=10^5~\mathrm{m/s}$, $\tau_0v_0=33~$nm, $I_0=60~\mu\text{A}/\mu\text{m}^2$.}
  \label{Sfigyc2}
\end{figure}

Finally, we discuss the influence of zero-point fluctuations of the Josephson phase mode. 
Focusing on the zero-temperature limit, as shown in Fig.~\ref{Sfigyc2}, increasing disorder (left panel) and reducing the Fermi velocity (right panel) both suppress the phase stiffness, leading to a reduction of the zero-temperature Josephson critical current $I_c(0)$ and the diode efficiency $\eta(0)$.

This behavior originates from the softening of the Josephson phase mode (at finite $q$) when the phase stiffness decreases. A weaker stiffness enlarges the available phase space of the collective-mode integration, thereby enhancing zero-point phase fluctuations. Through the Debye–Waller–type exponential renormalization factor, these enhanced quantum fluctuations suppress the effective Josephson coupling $\bar J_1(0)$ and $\bar J_2(0)$. As a consequence, both the zero-temperature critical current $I_c(0)$ and the diode efficiency $\eta(0)$ are reduced.

This zero-point effect suggests the intrinsic quantum sensitivity of low-dimensional Josephson-diode systems.
Importantly, because this suppression originates from quantum zero-point fluctuations rather than thermal effects, it represents a fundamental quantum limitation that cannot be circumvented by cooling. It indicates that enhancing phase stiffness (through improved material quality, reduced disorder, or optimized carrier density) is essential for robust device functionality and for maximizing nonreciprocal Josephson performance, particularly in low-dimensional systems.

%